\documentclass[useAMS,usenatbib,epsfig]{mn2e}
\newif\ifAMStwofonts
\AMStwofontstrue
\usepackage{amsmath,amsfonts,epsfig,natbib}
\usepackage{amssymb}
\usepackage{color}

\def\reff@jnl#1{{\rm#1\/}}
\def\aj{\reff@jnl{AJ}}                  
\def\araa{\reff@jnl{ARA\&A}}            
\def\apj{\reff@jnl{ApJ}}                
\def\apjl{\reff@jnl{ApJ}}               
\def\apjs{\reff@jnl{ApJS}}              
\def\ao{\reff@jnl{Appl.Optics}}         
\def\apss{\reff@jnl{Ap\&SS}}            
\def\aap{\reff@jnl{A\&A}}               
\def\aapr{\reff@jnl{A\&A~Rev.}}         
\def\aaps{\reff@jnl{A\&AS}}             
\def\azh{\reff@jnl{AZh}}                
\def\baas{\reff@jnl{BAAS}}              
\def\gca{\reff@jnl{GeCoA}}              
\def\jrasc{\reff@jnl{JRASC}}            
\def\memras{\reff@jnl{MmRAS}}           
\def\mnras{\reff@jnl{MNRAS}}            
\def\pra{\reff@jnl{Phys.Rev.A}}         
\def\prb{\reff@jnl{Phys.Rev.B}}         
\def\prc{\reff@jnl{Phys.Rev.C}}         
\def\prd{\reff@jnl{Phys.Rev.D}}         
\def\prl{\reff@jnl{Phys.Rev.Lett}}      
\def\pasp{\reff@jnl{PASP}}              
\def\pasj{\reff@jnl{PASJ}}              
\def\qjras{\reff@jnl{QJRAS}}            
\def\skytel{\reff@jnl{S\&T}}            
\def\solphys{\reff@jnl{Solar~Phys.}}    
\def\sovast{\reff@jnl{Soviet~Ast.}}     
\def\ssr{\reff@jnl{Space~Sci.Rev.}}     
\def\zap{\reff@jnl{ZAp}}                
\def\nat{\reff@jnl{Nature}}             

\newcommand{\wmap}{{\it WMAP }}
\newcommand{\planck}{{\it Planck }}
\newcommand{\healpix}{{\tt HEALPix }}
\newcommand{\commander}{{\tt Commander }}
\def\lesssim{\mathrel{\hbox{\rlap{\hbox{\lower4pt\hbox{$\sim$}}}\hbox{$<$}}}}

\title[QUIJOTE study of W43, W44 and W47]
{QUIJOTE Scientific Results. II. Polarisation Measurements of the Microwave Emission in the Galactic molecular complexes W43 and W47 and supernova remnant W44}

\author[R. G\'enova-Santos et al.] {R. G\'enova-Santos$^{1,2}\thanks{E-mail: rgs@iac.es}$, J.~A. Rubi\~no-Mart\'{\i}n$^{1,2}$,  A. Pel\'aez-Santos$^{1,2}$, F. Poidevin$^{1,2}$, 
\newauthor R. Rebolo$^{1,2,3}$, R. Vignaga$^{1,2}$, E. Artal$^4$, S. Harper$^{5}$, R. Hoyland$^1$, 
\newauthor A. Lasenby$^{6,7}$, E. Mart\'{\i}nez-Gonz\'alez$^{8}$, L. Piccirillo$^{5}$, D. Tramonte$^{1,2}$, 
\newauthor and  R.~A. Watson$^{5}$ 
\\
$^1$ Instituto de Astrofis\'{i}ca de Canarias, 38200 La Laguna, Tenerife, Canary Islands, Spain\\
$^2$ Departamento de Astrof\'{\i}sica, Universidad de La Laguna (ULL), 38206 La Laguna, Tenerife, Spain\\
$^3$ Consejo Superior de Investigaciones Cient\'{\i}ficas, Spain\\
$^4$ Departamento de Ingenieria de COMunicaciones (DICOM), Laboratorios de I+D de Telecomunicaciones,\\ Plaza de la Ciencia s/n, E-39005 Santander, Spain\\
$^5$ Jodrell Bank Centre for Astrophysics, Alan Turing Building, School of Physics and Astronomy, The University of Manchester,\\
Oxford Road, Manchester, M13 9PL, U.K\\
$^6$ Astrophysics Group, Cavendish Laboratory, University of Cambridge, J.J. Thomson Avenue, Cambridge CB3 0HE, UK\\
$^7$ Kavli Institute for Cosmology, Madingley Road, Cambridge, CB3 0HA\\
$^8$ Instituto de F\'{\i}sica de Cantabria (CSIC-Universidad de Cantabria), Avda. de los Castros s/n, 39005 Santander, Spain\\
}
\date{Accepted Received In original form}

\pagerange{\pageref{firstpage}--\pageref{lastpage}}
\pubyear{2014}
\begin{document}

\label{firstpage}
\maketitle

\begin{abstract}
We present Q-U-I JOint TEnerife (QUIJOTE) intensity and polarisation maps at $10-20$~GHz covering a region along the Galactic plane $24^\circ\lesssim l \lesssim 45^\circ$, $|b|\lesssim 8^\circ$. These maps result from $210$~h of data, have a sensitivity in polarisation of $\approx$40~$\mu$K~beam$^{-1}$ and an angular resolution of $\approx 1^\circ$. Our intensity data are crucial to confirm the presence of anomalous microwave emission (AME) towards the two molecular complexes W43 ($22\sigma$) and W47 ($8\sigma$). We also detect at high significance ($6\sigma$) AME associated with W44, the first clear detection of this emission towards a SNR. The new QUIJOTE polarisation data, in combination with {\it WMAP}, are essential to: i) Determine the spectral index of the synchrotron emission in W44, $\beta_{\rm sync}=-0.62\pm 0.03$, in good agreement with the value inferred from the intensity spectrum once a free-free component is included in the fit. ii) Trace the change in the polarisation angle associated with Faraday rotation in the direction of W44 with rotation measure $-404\pm 49$~rad~m$^{-2}$. And iii) set upper limits on the polarisation of W43 of $\Pi_{\rm AME} <0.39$ per cent (95 per cent C.L.) from QUIJOTE 17~GHz, and $<0.22$ per cent from \wmap 41~GHz data, which are the most stringent constraints ever obtained on the polarisation fraction of the AME. For typical physical conditions (grain temperature and magnetic field strengths), and in the case of perfect alignment between the grains and the magnetic field, the models of electric or magnetic dipole emissions predict higher polarisation fractions.
\end{abstract}


\section{Introduction}

The study and characterisation of polarised Galactic foregrounds in the microwave and sub-mm ranges is becoming increasingly important now that experiments \citep{bicep216} are starting to put increasingly tighter constraints on the inflationary B-mode anisotropy in the cosmic microwave background (CMB) polarisation \citep{kamionkowski97,zaldarriaga97}. There are two Galactic foregrounds that are known to emit strong linearly polarised radiation: the synchroron radiation resulting from cosmic-ray (CR) electrons accelerated by the Galactic magnetic field, and the thermal radiation originated in Galactic interstellar dust. Both are known to have polarisation fractions of up to $20\%$ in some regions of the sky \citep{kogut07,pip19}. In total intensity there are two other foregrounds that show up in the microwave range: the free-free emission, and the so-called anomalous microwave emission (AME). While the former is very well characterised and is known to have practically zero polarisation \citep{trujillo02}, very little is known about the polarisation properties of the AME. 

Over the last decade, a wide variety of observations \citep{watson05,dickinson09,tibbs10,genova11,per20,pip15,battistelli15} have helped to establish the electric dipole radiation from very small and fast rotating interstellar dust grains \citep{draine98,ali09,hoang10,ysard10,silsbee11}, commonly referred to as {\it spinning dust emission}, as the most probable physical mechanism responsible for AME. An alternative scenario based on magnetic dipole emission \citep{draine99,draine13} have also been proposed. Different theoretical studies \citep{draine99,lazarian00,draine13,hoang13,hoang16} have provided predictions for the polarisation spectra of these two mechanisms, always setting the polarisation fraction above $\sim 1$\% at frequencies below $\approx 20$~GHz, when typical physical conditions, grain sizes and magnetic field strengths are assumed. However, from the observational standpoint, so far no clear detection of AME polarisation has been claimed. After a marginal detection, $\Pi=3.4^{+1.5}_{-1.9}\%$, at 11~GHz with the COSMOlogical Structures On Medium Angular Scales (COSMOSOMAS) experiment \citep{battistelli06}, all other observations \citep{casassus08,mason09,lopez11,dickinson11,genova15} have set upper limits between 1 and 6\% in the frequency range 10 to 40~GHz, where the AME is more prominent in intensity. An exhaustive review of these results can be found in \citet{rubino12b}.

In this context, it is important to undertake microwave and sub-mm surveys covering sky areas as wide as possible, even if at the cost of a poorer angular resolution. This strategy seems most appropriate to search for the B-mode signal from the reionisation bump, that shows up at large angular scales, and avoids contamination from the finer gravitational lensing-induced B-mode anisotropies. The \planck satellite \citep{cpp2015-1} has obtained full-sky polarisation maps covering the milimetre range up to 353~GHz, and providing accurate measurements of the polarisation properties of the thermal dust emission. This information could be used to correct CMB maps at lower frequencies. However, the \planck survey must be complemented by similar data at low frequencies that could give information of the polarisation properties of the synchrotron and also of the AME \citep{krachmalnicoff16}. The \wmap satellite lowest frequency is 22.7~GHz \citep{kogut07}. Ground-based experiments like the Cosmology Large Angular Scale Surveyor (CLASS; Watts et al. 2015) have receivers down to 40~GHz, while the C-Band All Sky Survey (C-BASS; Irfan et al. 2015) will cover the full sky at 5~GHz. Q-U-I JOint TEnerife (QUIJOTE; G\'enova-Santos et al. 2015b) is benefited from having 6 frequency bands between 10 and 40~GHz, and therefore will provide very valuable information about the synchrotron and AME polarisations.

This is the second of a series of scientific QUIJOTE papers. We present maps at 11, 13, 17 and 19~GHz of a region of the Galactic plane between $l\approx 24^\circ$ and $l\approx 45^\circ$, and extending in Galactic latitude up to $|b|\approx 8^\circ$. These QUIJOTE maps show diffuse polarised emission distributed along the Galactic plane. Using these data in combination with other ancillary data, including \wmap and {\it Planck}, we study the spectral properties of the diffuse emission, and also of the more compact emission towards two molecular complexes W43 and W47 and towards the supernova remnant (SNR) W44. In W43 we get the most stringent limits to date on the AME polarisation. 

The paper is organised as follows: in section~\ref{sec:w43_w44_w47} we present a physical description of the studied compact regions based on information extracted from the literature. In section~\ref{sec:data} we describe the data used, while in section~\ref{sec:maps} we present the maps. The spectral properties of the diffuse emission are analysed, in intensity and in polarisation, in section~\ref{sec:diffuse}. In section~\ref{sec:intensity} we study the spectral energy distributions (SEDs) in total intensity of W43, W44 and W47, and fit them with models that include a spinning dust component. In section~\ref{sec:polarisation} we set constraints on the AME polarisation in W43 and study the polarisation of the synchrotron emission in W44. Finally, the main conclusions of this work are summarised in section~\ref{sec:conclusions}.

\section{The W43, W44 and W47 regions}\label{sec:w43_w44_w47}
In this section we discuss the main physical characteristics of the three compact regions that we analyse in detail in this paper.

\subsection{The molecular complexes W43 and W47}
With a total mass of $\sim 7\times 10^6~M_\odot$, and physical size of $\sim 140$~pc (it extends on the sky from $l=29.6^\circ$ to $l=31.4^\circ$ and from $b=-0.5^\circ$ to $b=0.3^\circ$), W43 is considered to be one of the most extreme molecular complexes in our Galaxy \citep{nguyen11}. It is located at a distance of $\approx 5.5$~kpc, at the meeting point of the Scutum-Centaurus Galactic arm and the bar. \citet{nguyen11} conclude that W43 is a coherent molecular and star-forming complex, encompassing more than 20 molecular clouds with high velocity dispersions, and surrounded by atomic gas, which extends $\sim 290$~pc. It is considered to be a coherent and gravitationally-bound ensemble of clouds \citep{motte14}. Using CO observations, \citet{solomon87} distinguished individual clouds within W43, and located the two most massive ones, so-called W43-main and W43-south, which have virial masses of several times $10^6~M_\odot$. More recently, \citet{rathborne09} identified more than 20 molecular clouds using $^{13}$CO observations. The core of W43-main harbours a well-known giant HII region powered by a particularly luminous cluster of Wolf-Rayet and OB stars \citep{blum99}. \citet{motte03} presented observations with higher-spatial resolution at $1.3$~mm and $350~\mu$m that revealed W43-main to be a complex structure of chimneys and filaments forming a ``mini-starburst''. This is one of the most luminous star-forming complexes in the Galaxy. They identified a filamentary structure with 51 compact fragments with masses $40-4000~M_\odot$, most of them being protocluster candidates. The most-massive ($>100~M_\odot$) of these protoclusters are potentially sites of ongoing or future massive star formation. These findings were later confirmed by \citet{bally10} using FIR to sub-mm data from the {\it Herschel} Space Observatory.

There is not much information in the literature about W47. In their catalogue of HII regions \citet{paladini03} associated 7 compact regions with W47. At 1.4~GHz these objects have fluxes in the range $6-18$~Jy, and angular sizes $6-9$~arcmin. In a radio recombination line (RRL) survey at 9~GHz \citet{bania12} identified emission from the compact HII region G037.468-0.105, which has coordinates coincident with W47. They estimate a kinematic distance of $9.6\pm 0.5$~kpc for this object. Finally, \citet{stil03} discussed the presence of a chimney, formed by a filament of HI emission extending north of W47, which could be the result of a expanding bubble that originates in this HII region.

\subsection{The supernova remnant W44}
W44 is a middle-aged (20,000 years old) bright SNR with a size of $\sim 0.5^\circ$, and a mixed morphology characterised by a bright non-thermal shell-like radio structure and centrally-concentrated thermal X-ray emission \citep{rho98}. It lies on the Galactic plane at $(l,b)$=$(34.7^\circ,-0.4^\circ)$ at a distance of $\sim 3.1$~kpc, and is probably located in the Sagittarius arm \citep{cardillo14}. According to \citet{seta98} there are six giant molecular clouds that appear to be surrounding the remnant, some of which seem to be partially interacting with the SNR on its southeastern and western sides. In fact, this object constitutes one of the few cases of an interaction between a SNR and a molecular cloud. Castelletti et al. 2007 (hereafter C07) presented VLA observations of this object which they used, in combination with previous observations between 0.022~GHz and 10.7~GHz, to infer an integrated synchrotron spectral index of $\beta_{\rm sync}=-0.37\pm 0.02$. They also produced an spectral index map, that show internal filaments with values of $\sim -0.5$, but also external regions with a flattening of the index that they interpret as the result of the interaction with molecular clouds.

W44 has also awaken interest due the characteristics of its $\gamma$-ray emission. Data from the AGILE satellite revealed, for the first time in any SNR, $\gamma$-ray emission below 200~MeV in W44 \citep{giuliani11}. These observations in this low energy range are important as they permit to disentangle leptonic emission, namely bremsstrahlung or inverse-Compton, from hadronic emission due to the decay of neutral pions originated in CRs interactions. The pion bump was in fact detected in Fermi-LAT data \citep{ackermann13}. \citet{cardillo14} studied the combined $\gamma$-ray spectrum from AGILE and Fermi-LAT and concluded that no model based on leptonic emission only can jointly explain these data and the radio data of C07. On the contrary, they found that the multi-wavelength spectrum from radio to $\gamma$-rays can be explained if the $\gamma$-ray emission is dominated by hadronic processes with a broken power-law spectrum. More recently \citet{cardillo16} considered an alternative model to reproduce the observed $\gamma$-ray spectrum based in the re-acceleration and compression of Galactic CRs, with no need of introducing a break in the proton energy distribution.

\section{Data}\label{sec:data}

\subsection{QUIJOTE data}\label{sec:quijote_data}
The new data presented in this article were acquired with the QUIJOTE experiment, a collaborative project consisting of two telescopes and three polarimeter instruments covering the frequency range from 10 to 40~GHz. The main science driver of this experiment is to constrain or to detect the B-mode anisotropy in the CMB polarisation down to a tensor-to-scalar ratio of $r=0.05$, and to characterise the polarisation of the low-frequency foregrounds, mainly the synchrotron and the AME. A more detailed description of the technical and scientific aspects of this project can be found in various conference proceedings \citep{rubino12,genova15b}, or in Rebolo et al. (in preparation). The data used in this work come from the first instrument of QUIJOTE, the so-called multi-frequency instrument (MFI). It consists of 4 horns, which provide 8 independent maps of the sky intensity and polarisation, in four frequency bands centred at $11$, $13$, $17$ and $19$~GHz (each frequency is covered by two independent horns), each with a 2~GHz bandwidth, and with angular resolutions of FWHM=$52$~arcmin (for the $11$~GHz and $13$~GHz bands) and FWHM=$38$~arcmin (for the $17$~GHz and $19$~GHz bands). Data from this instrument were used for the first time in a recent publication \citep{genova15}, where we analysed the properties, in intensity and in polarisation, of the AME towards the Perseus molecular complex. 

\subsubsection{Observations}
The observations used in this work were carried out between March and June 2015 using the MFI. Initially, they consisted in raster scans in local coordinates centred on W44. We performed these by driving the telescope in azimuth $11^\circ/{\rm cos(EL)}$ at a constant elevation and at a velocity of $1^\circ/{\rm cos(EL)}$, and stepping the elevation by $0.1^\circ$ after each scan. Each of these observations takes around 25~min, and produces a map of the sky of around $11^\circ\times 11^\circ$. It must be noted at this point that the MFI horns point to sky positions separated up to $5^\circ$, so each of them rasters a slightly different sky patch. Therefore, the total sky area surveyed in each of these observations is slightly wider, $\sim 18^\circ\times 18^\circ$, at the expense of the common area between all horns being only $\sim 5^\circ\times 5^\circ$. In order to cover a wider sky area, extending to higher Galactic longitudes, since the beginning of May 2015 we started observing in a different mode, consisting in $\sim 150$ raster scans at a constant elevation extending in azimuth $22/{\rm cos(EL)}^\circ$. Around half of the observations were performed in this mode. 

The final observing time was 210~h. However, these data were carefully inspected by eye, and periods affected by bad weather, strong gain variations, interference, or contamination by geostationary satellites, were removed and not used in this study. We surveyed a region close to declination zero (W44 is at $\delta=1.37^\circ$), so geostationary satellites, which are distributed along the equatorial plane and emit predominantly between 10 and 13~GHz, are a major concern. In our data processing pipeline we follow the practice to excise all data less than $5^\circ$ from any satellite. In the present analysis, in order to recover the wider sky area possible, we relaxed this requirement and used a distance of $3.5^\circ$ for flagging. After flagging, the final effective observing time per horn was 48, 110, 30 and 111~h respectively in horns 1, 2, 3 and 4; horns 1 and 3 provide the 11 and 13~GHz bands, and are the most severely affected by satellite interference. The total sky area covered by each horn was respectively 203, 344, 174 and 448~deg$^2$. In this work, intensity data from all horns will be included in the analysis. However, in polarisation horns 2 and 3 are the most sensitive, the better calibrated and characterised, so we will not use in this work polarisation data from horns 1 and 4.

\subsubsection{Calibration}
As it was explained in \citet{genova15}, our amplitude calibrator is Cas A, which is observed at least once per day. We use the spectrum derived in \citet{weiland11}, and integrate it in the MFI passbands to derive the reference flux densities in each channel. The decrease in the flux of Cas A due to its secular variation is accounted for using the model presented in \citet{hafez08}. We calibrated each individual channel separately, so a very accurate determination of the gain factors is necessary in order to avoid any possible leakage from intensity to polarisation. A leakage could arise when subtracting pairs of channels measuring the two orthogonal polarisations if they were not perfectly balanced. We measured gain factors for 165 individual Cas A observations from June 2014 to April 2015, and verified that the scatter of them is typically less than 4\%, and on average around 3\%. We then calculated the median of the gain factors for each channel, and used these values to calibrate each of our obervations. We have verified that this is an optimal calibration strategy, which renders precise flux densities in total intensity and no detectable leakage in polarisation, as it will be seen later. In fact, as a cross-check, we also estimated the gain factors using Tau A as calibrator, and replicating the same strategy applied to Cas A, and the values differ by less than 3\%.

Our calibrator for the polarisation angle is Tau A (also known as Crab nebula), which is observed at least once per day. We use the Tau A polarisation direction measured by \wmap at 22.7~GHz \citep{weiland11} under the assumption that it remains constant with frequency down to 11~GHz (it varies less than 5\% in the \wmap frequency range, from 22.7 to 93.0~GHz). As we have checked that the derived reference polarisation angle remains constant over time (at least within the scatter of the measurements, which is less than $1^\circ$), we combine hundreds of observations of Tau A to derive the final reference values for each horn. The accuracy of these values is between $0.4^\circ$ and $0.8^\circ$, depending on the horn.

\subsubsection{Map making}
Our map-making is based on a destriper algorithm, which notably reduces the effects of the $1/f$ noise in the data. This is much more important for intensity, where the knee frequencies of our receivers are $f_{\rm k}\sim 10-40$~Hz, much higher than for polarisation, where $f_{\rm k}\sim 0.1-0.2$~Hz. The code approximates the low-frequency noise part in the time-ordered data (TOD) as a series of offset functions of $2.5$~s length, which are then subtracted from the TOD. Our implementation follows the same equations as in the MADAM code \citep{keihanen05,kurki-suonio09}, with two main differences. First, we do not use prior information on the offset function amplitude distribution. And second, intensity and polarisation maps are reconstructed separately, as the sum and difference of correlated channels in QUIJOTE data provide a separated measurement of the intensity and polarisation signals, respectively. A detailed explanation about how QUIJOTE recovers the Stokes $I$, $Q$ and $U$ values from the measured signal is given in \citet{genova15}. Here we will use the second of the methods described there, which is based on an analytical $\chi^2$ minimisation. 

To produce the final maps we use a \healpix pixelisation \citep{gorski05} with $N_{\rm side}$ = 512 (pixel size $6.9$~arcmin), which is sufficient given the beam FWHM. All the analysis in this work will be performed on maps convolved at a common resolution of 1 degree. In order to avoid mask effects, we perform the convolution of QUIJOTE maps in the real space (the differences between these maps and those convolved in the Fourier space are typically under 1\% at distances larger than $\sim 0.5^\circ$ from the mask border).

\subsection{Ancillary data}\label{sec:ancillary_data}
In the low-frequency range we use intensity radio maps in order to better characterize the level of free-free and synchrotron emissions. In particular, we use the \citet{haslam82} map at 0.408~GHz, the \citet{berkhuijsen72} map at 0.820~GHz from the Dwingeloo radio telescope, the \citet{reich86} map at 1.42~GHz from the Stockert 25-m telescope, and the southern-sky \citet{jonas98} map at 2.326~GHz from the HartRAO telescope. The calibration of these maps is usually referred to the full-beam solid angle. Due to the presence of sidelobes, this calibration would produce underestimates on the flux densities of sources which are small compared with the main beam, so a correction is needed to scale the maps to the main-beam scale. This issue is particularly important in the \citet{reich86} and \citet{jonas98} maps, so we apply the correction factors of $1.55$ and $1.45$, derived respectively by \citet{reich88} and \citet{jonas98}. W43 ($\approx 60$~arcmin across) and W44 ($\approx 30$~arcmin) have both angular extents larger than the HartRAO main beam FWHM ($\theta_{\rm FWHM}=20$~arcmin), so in these cases a lower correction factor of $1.2$ has been used, which has been inferred from Fig.~6 of \citet{jonas98}. These maps have also important uncertainties related with the zero-level of the temperature scale. However, this is not a concern in our analyses as we always subtract the background level when extracting flux densities. 

C-BASS data at 4.76~GHz are not public, but we will adopt in our analyses the flux densities quoted in \citet{irfan15} for W43, W44 and W47. In polarisation, we use the \citet{wolleben06} maps of $Q$ and $U$ parameters at 1.4~GHz produced with the 25.6~m DRAO telescope.  The \citet{haslam82}, \citet{reich86} and \citet{jonas98} maps have been taken from \citet{platania03}. Recently \citet{remazeilles15} have delivered an improved destriped version of the \citet{haslam82} map. We have checked that the difference between flux densities extracted from this map and from the \citet{platania03} version differ by $\approx 5$\%, but we prefer to use the later map as it produces flux densities in W43, W44 and W47 closer to those presented in \citet{irfan15}, with which we will be comparing. The intensity data from \citet{berkhuijsen72} and the polarisation $Q$ and $U$ data from \citet{wolleben06} have been downloaded from the MPIfR's Survey Sampler\footnote{{\tt http://www.mpifr-bonn.mpg.de/survey.html}}, and projected into \healpix pixelisation.

\begin{figure*}
\centering
\includegraphics[width=17cm]{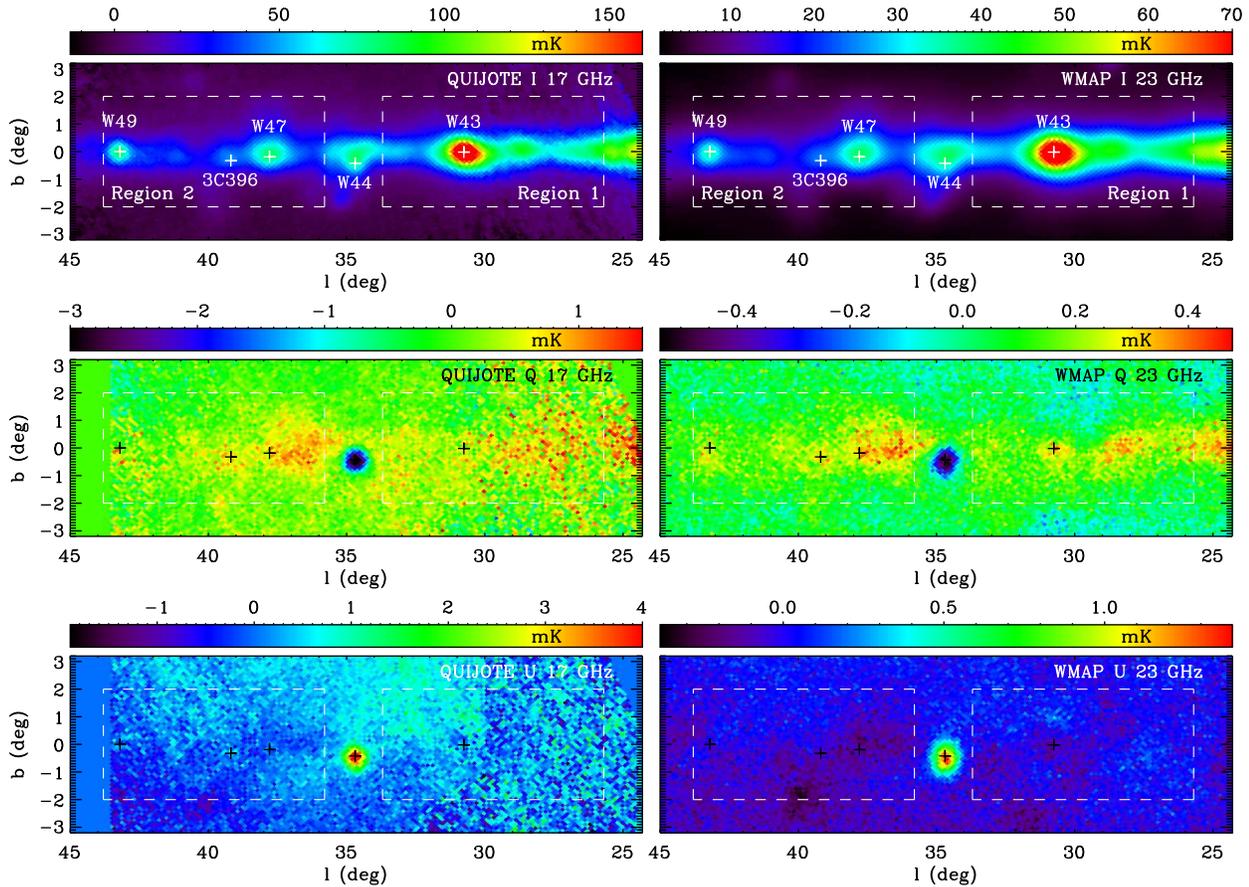}
\caption{\small QUIJOTE maps at 17~GHz on the region covered by the observations (left) in comparison with \wmap maps at 23~GHz (right). The first line displays intensity maps, and the second and third lines polarisation $Q$ and $U$ parameters, respectively. The maps show a fraction of the total area surveyed. The whole QUIJOTE maps encompass around 250~deg$^2$, corresponding in total to 110~h of observations, with RMS $\approx 150~\mu$K/beam in $I$ and $\approx 40~\mu$K/beam in $Q$ and $U$. The positions of the two regions analysed in this work, W44 and W47, are indicated, together with other objects along the Galactic plane. The polarisation of the SNR W44 is clearly detected both in $Q$ and $U$, as well as some diffuse Galactic polarisation around it. The two boxes enclosed by dashed white lines show the regions whose spectral properties are analysed in section~\ref{sec:diffuse}.}
\label{fig:large_maps}
\end{figure*}

In the microwave range, we also use intensity and polarisation \wmap and \planck data. We downloaded from the LAMBDA database\footnote{Legacy Archive for Microwave Background Data Analysis, {\tt http://lambda.gsfc.nasa.gov/}.} $I$, $Q$ and $U$ maps from the 9-year release of the \wmap satellite \citep{bennett13} to provide flux densities at frequencies 23, 33, 41, 61 and 94~GHz. Recent data from the second release of the \planck mission\footnote{Downloaded from the \planck Legacy Archive (PLA) {\tt http://pla.esac.esa.int/pla/}.} \citep{cpp2015-1} cover the frequencies 28, 44, 70, 100, 143, 217, 353, 545 and 857~GHz. We also downloaded the released Type 1 CO maps \citep{cpp2013-13}, which are used to correct the 100, 217 and 353~GHz intensity maps from the contamination introduced by the CO rotational transition lines (1-0), (2-1) and (3-2), respectively, and \planck component-separated maps (free-free, synchrotron and AME) obtained with the \commander component-separation tool \citep{cpp2015-10}. Although \planck measures polarisation in its seven frequencies up to 353~GHz, only polarisation maps for the three LFI frequencies (28, 44 and 70~GHz) and for the 353~GHz HFI frequency have been made public in the second release. We use these data, after correcting the three LFI maps from the bandpass mismatch, that produces intensity to polarisation leakage, using the correction maps also provided in the \planck Legacy Archive (PLA). 

In the far-infrared spectral range we use Zodi-Subtracted Mission Average (ZSMA) COBE-DIRBE maps \citep{hauser98} at 240~$\mu$m (1249~GHz), 140~$\mu$m (2141~GHz) and 100~$\mu$m (2998~GHz), which complement \planck data to constrain the spectrum of the thermal dust emission.

Finally, we use two different templates to help determining the level of free-free emission. The first is obtained from maps of the emission measure and electron temperature derived from the \planck data using {\tt Commander}, and available in the PLA. The second is a map of the free-free emission at 1.4~GHz produced by \citet{alves12}\footnote{This map was downloaded from {\tt http://www.jodrellbank.manchester.ac.uk/research/\\parkes\_rrl\_survey/}} using radio recombination line data from the HI Parkes All-Sky Survey (HIPASS).

Except the \citet{berkhuijsen72} and \citet{wolleben06} maps (which were re-projected, as it was said before), and the \citet{alves12} template (which is used at its original pixelisation), all these data are given in \healpix format. We use all maps at a $N_{\rm side}$=512 pixelisation, and at a common angular resolution of 1~degree. The \wmap data are available at LAMBDA at this angular resolution. The exact beam profiles were used to convolve these maps. For the rest of the maps, we consider Gaussian beams and use each telescope-beam FWHM to convolve them to an angular resolution of 1~degree. This is particularly important for the LFI polarisation maps, as the leakage correction maps are given at this angular resolution, and are not reliable at smaller angular scales.

\section{Maps}\label{sec:maps}

The final intensity and polarisation\footnote{For the $Q$ and $U$ polarisation maps in this article we use the COSMO convention ({\tt http://healpix.sourceforge.net/html/\\intronode6.htm}), which have become common practice in CMB studies. {\it WMAP} and \planck polarisation maps follow this convention, and we do the same in QUIJOTE. The net effect with respect to the IAU convention \citep{hamaker96} is a change in the sign of $U$, i.e. $U=-U_{\rm IAU}$. However, our definition of the polarisation angle, $\gamma=0.5\times{\rm tan}^{-1}(-U/Q)$, is the same in the IAU convention, $\gamma_{\rm IAU}=0.5\times{\rm tan}^{-1}(U_{\rm IAU}/Q)$, with $\gamma=\gamma_{\rm IAU}$ being positive for Galactic north through east.} QUIJOTE maps at 17~GHz, in a region covering Galactic longitudes between 25$^\circ$ and 45$^\circ$, are shown in Fig.~\ref{fig:large_maps}, in comparison with the \wmap 9-year maps at 23~GHz. Although for consistency we extract flux densities in maps convolved to a common resolution of $1^\circ$, here maps are displayed at their original angular resolutions. In intensity, the three regions that are the focus of this work, the SNR W44 and the molecular complexes W43 and W47, are clearly detected, as well as the molecular cloud W49 and, although at a lower signal-to-noise, the fainter SNR 3C396. In polarisation, the synchrotron emission from W44 is clearly visible, and presents similar polarisation direction in QUIJOTE and {\it WMAP}, with negative $Q$ and positive $U$. Some diffuse polarised emission along the Galactic plane is also visible in both surveys, with the typical polarisation direction in this position, $Q$ positive and $U$ close to zero. No excess of polarisation is seen towards W43 nor W47, as it is expected since their intensities are dominated by free-free emission, which is known to be (practically) unpolarised.

\begin{figure*}
\centering
\includegraphics[width=18cm]{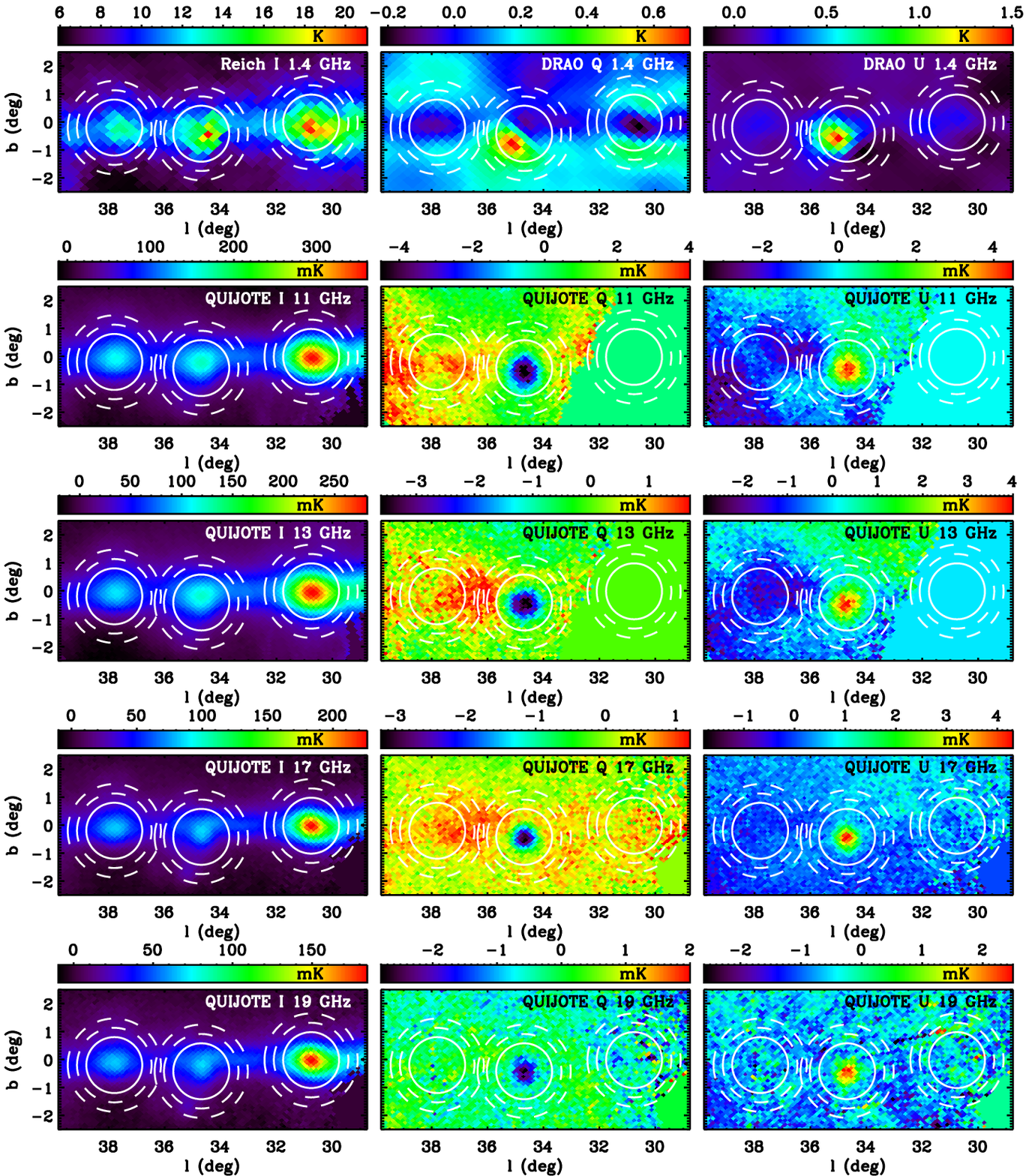}
\caption{\small}
\label{fig:iqu_maps}
\end{figure*}
\begin{figure*}
\centering
\includegraphics[width=18cm]{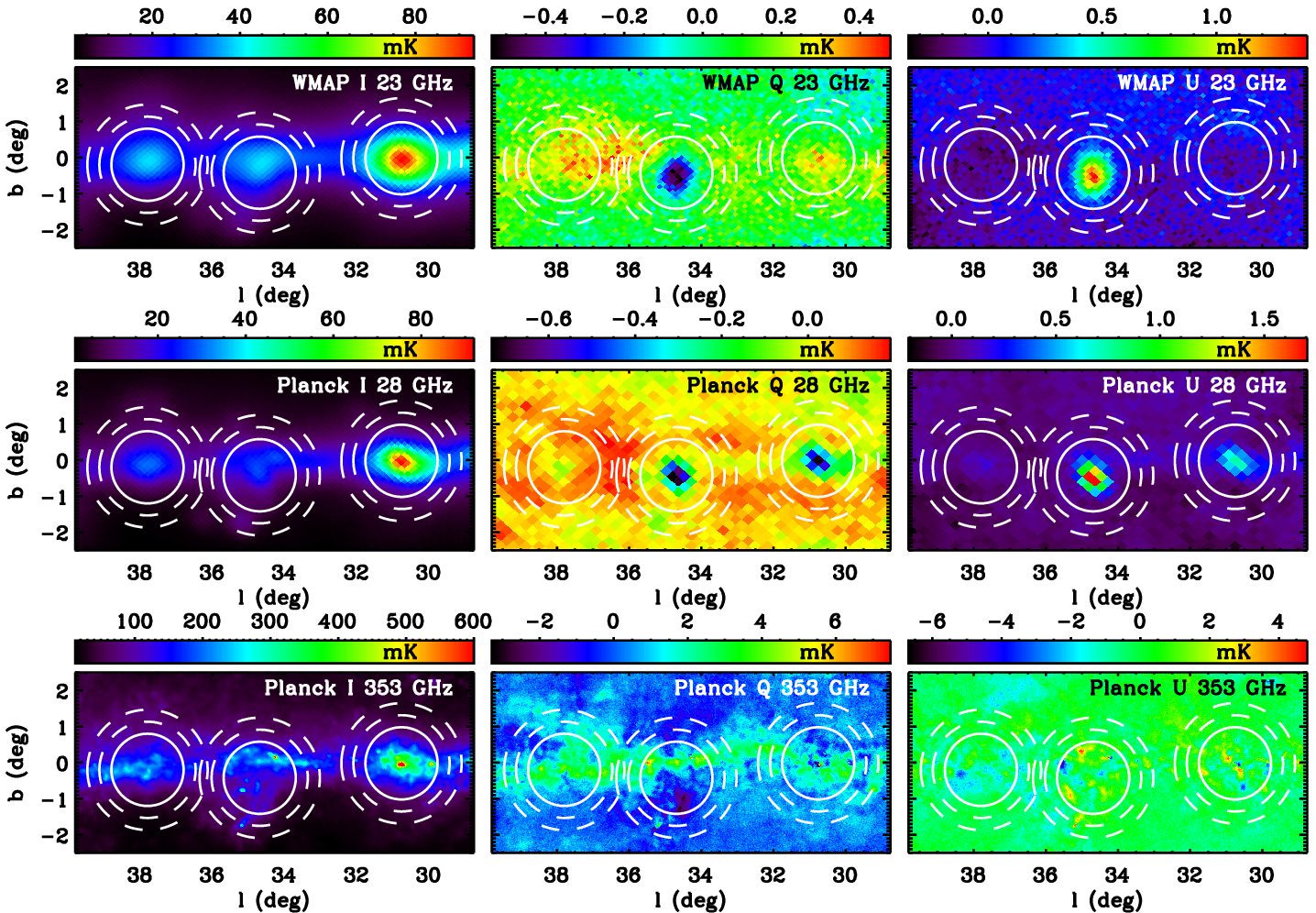}
\contcaption{\small Intensity (left) and polarisation ($Q$ middle, $U$ right) maps around the regions W43, W44 and W47, derived from the QUIJOTE observations and from other ancillary data. QUIJOTE maps are represented only for the horns 2 and 3. Similar maps, at the same four frequencies, are provided by horns 1 and 4. The solid circles show the apertures used for flux integration in W47, W44 and W43 (from left to right), whereas the dashed contours limits the extent of the ring used for background subtraction. The polarisation on the SNR W44 is clearly detected in all frequencies, whereas no signal is seen towards the HII regions W43 and W47, as expected due to their emissions being dominated by free-free emission. Some diffuse Galactic polarisation is also detected along the plane, mainly in $Q$.}
\end{figure*}

More detailed $I$, $Q$ and $U$ maps of the three regions that will be studied here, and covering more frequencies, are shown in Fig.~\ref{fig:iqu_maps}. The three circles represent the apertures that we use to integrate flux densities in section~\ref{sec:intensity}, and the two concentric dashed circles the regions where we calculate the background level. From left to right the circles correspond respectively to W47, W44 and W43. We show the four QUIJOTE frequency maps from horns 2 and 3. We have produced other four maps at the same frequencies using data from horns 1 and 4, that will also be used to derive flux densities only in total intensity. The mask that is seen in the polarisation maps at 11 and 13~GHz results from the removal of data affected by contamination from geostationary satellites. These satellites are distributed along the stripe of declination $\delta=0^\circ$, and therefore affect mainly W43 ($\delta=-1.9^\circ$). Although polarisation data are more severely affected, and we therefore lack polarisation information for W43, by a careful masking we managed to keep some useful data in total intensity for this region. The W44 polarisation direction at 1.4~GHz in the DRAO map\footnote{Note that originally the DRAO data followed the IAU convention for the definition of $Q$ and $U$. For consistency with the rest of the data, and with the analyses presented in this paper, which follow the COSMO convention, we have changed the sign of the $U$ data of this map, $U=-U_{\rm IAU}$.} is different from other frequencies, which could be due to Faraday rotation on the SNR. Note also that there is an offset on the position of the source, which could be due to depolarisation in the Galactic plane. Both effects will be discussed in detail in section~\ref{sec:polarisation}. The structure of the intensity and polarisation maps remains very similar from 11 to 23~GHz. In \planck 28~GHz there is clear evidence of leakage from intensity to polarisation, mainly produced by bandpass mismatch \citep{cpp2015-2}, in particular at the position of W43. The correction maps, that we have applied, are only reliable on angular scales larger than $\sim 1^\circ$, and this could explain the presence of intensity to polarisation leakage at this position. At 353~GHz no clear polarisation signal is seen towards W44, except maybe a negative feature in $Q$ that appears towards the south. However, these maps do not seem reliable as they also show some possible intensity leakage in $Q$ along the Galactic plane.

As a consistency check for the QUIJOTE data, we have performed jackknife tests. We have uniformly split the data in two halves in such a way that the two resulting maps have as similar sky coverage as possible. The subtraction of the two parts consistently cancels out the sky signal, and has a temperature distribution which is consistent with instrumental noise. In each part, as it contains only half of the data, the noise is degraded by a factor $\sqrt{2}$. Another factor $\sqrt{2}$ comes from the subtraction of the two parts. Therefore, we divided the resulting map by $2$ in order to make it representative of the true instrumental noise in the total map. In Table~\ref{tab:jk} we show the RMS noises (units: $\mu$K per beam of one degree\footnote{Note that the actual QUIJOTE beam widths are $\theta_{\rm FWHM}\approx 0.89^\circ$ for 11 and 13~GHz, and  $\theta_{\rm FWHM}\approx 0.65^\circ$ for 17 and 19~GHz,  but as in this work we have convolved all maps to $1^\circ$, here we quote noises at this angular scale.}) calculated in these maps, and in the original maps, in a circular region of $1^\circ$ radius centred in $(l,b)=(36^\circ,3^\circ)$. Although we have selected a region away from the plane, and with relatively little sky signal, it can be seen that the sky still dominates the dispersion of the data in intensity. In polarisation the RMS in the total and in the JK maps are very similar, which indicates that the total map is dominated by instrumental noise. The noises in intensity are typically 3 to 7 times worse than in polarisation, due to the presence of $1/f$ residuals. In the maps from horn 1, the noises in polarisation are notably worse. The reason for this is that in this horn we cannot step the modulator angle, and this results in a worse recovery of the polarised signal due to having less independent polarisation directions. As it was said before, this horn, and horn 4, will not be considered in the polarisation analyses of this work. In the last column of Table~\ref{tab:jk} we quote the equivalent instantaneous sensitivities, which have been derived by multiplying the average of the $Q$ and $U$ noises by the square root of the integration time per beam. The amplitudes of the white noise in the spectra of the time-ordered-data range between $0.9$~mK~s$^{1/2}$ and $1.3$~mK~s$^{1/2}$, depending on the channel. At frequencies 11 to 17~GHz, $1/f$ residuals make the noises on the JK maps only slightly higher (by $15$ or $20\%$), so we can conclude that our polarisation maps are dominated by white noise. At 19~GHz the noise is considerably higher, in particular in horn 2, which is probably due to the atmospheric contribution through the 22~GHz water vapour line.

\begin{table*}
\begin{center}
\begin{tabular}{ccccccccccccc}
\hline\hline
\noalign{\smallskip}
Horn & Freq.   && \multicolumn{2}{c}{$\sigma_{\rm I}$ ($\mu$K/beam)}    && \multicolumn{2}{c}{$\sigma_{\rm Q}$ ($\mu$K/beam)}    && \multicolumn{2}{c}{$\sigma_{\rm U}$ ($\mu$K/beam)} && $\sigma_{\rm Q,U}$ (mK~s$^{1/2}$) \\
\noalign{\smallskip}
\cline{4-5}\cline{7-8}\cline{10-11}\cline{13-13}
\noalign{\smallskip}
& (GHz) && Map & JK && Map & JK && Map & JK && JK \\
\noalign{\smallskip}
\hline
\noalign{\smallskip}
1   &   11.2  &&  415    &   122     &&   119	  &   117   &&     82	&    84   &&   3.3 \\
\noalign{\smallskip}
1   &   12.8  &&  372    &    89     &&   102	  &   108   &&     70	&    70   &&   3.0  \\
\noalign{\smallskip}
2   &   16.7  &&  390    &   148     &&    29	  &    28   &&     37	&    38   &&   1.1  \\
\noalign{\smallskip}
2   &   18.7  &&  355    &   177     &&    55	  &    46   &&     68	&    77   &&   2.1  \\
\noalign{\smallskip}
3   &   11.1  &&  614    &   194     &&    53	  &    63   &&     46	&    53   &&   1.6  \\
\noalign{\smallskip}
3   &   12.9  &&  369    &   142     &&    46	  &    62   &&     46	&    50   &&   1.6  \\
\noalign{\smallskip}
4   &   17.0  &&  428    &   210     &&    43	  &    43   &&     38	&    39   &&   1.1  \\
\noalign{\smallskip}
4   &   19.0  &&  362    &   236     &&    50	  &    55   &&     52	&    47   &&   1.3  \\
\noalign{\smallskip}
\hline\hline
\end{tabular}
\normalsize
\caption{RMS per beam, in intensity and in polarisation, calculated on the QUIJOTE maps in a circular radius of $1^\circ$ around $(l,b)=(36^\circ,3^\circ)$. For each case ($I$, $Q$ and $U$) we show the RMS calculated in the original maps and in the difference of the two jackknife maps divided by two. The former should be representative of the combined background and instrumental noise uncertainties, whereas the later would indicate the level of instrumental noise only. In the last column we show the instrument instantaneous sensitivities (units: mK~s$^{1/2}$) in polarisation, which have been obtained by normalising the average $Q$ and $U$ noises in the jackknives by the integration time per beam.}
\label{tab:jk}
\end{center}
\end{table*}

\section{Diffuse Galactic emission}\label{sec:diffuse}

The maps of Fig.~\ref{fig:large_maps} show that the emission in total intensity is dominated by compact sources distributed along the Galactic plane, with an important contribution from the diffuse emission from the interstellar medium (ISM), whereas in polarisation only W44 has a significant emissivity, with the rest of the emission being predominantly diffuse. According to the component separation provided by {\tt Commander}, the diffuse emission in total intensity along the Galactic plane is predominantly free-free, with a relative contribution of $\sim 70\%$ at 22.7~GHz, while the contribution from the synchrotron increases at high Galactic latitudes and at lower frequencies. On the other hand, the diffuse polarised emission is basically synchrotron. The QUIJOTE and {\it WMAP} maps displayed in Figs.~\ref{fig:large_maps} and \ref{fig:iqu_maps} reveal a region with particularly strong diffuse polarisation towards the east of W44. This emission is bright and positive in $Q$ and close to zero in $U$, which implies polarisation direction perpendicular to the Galactic plane. This is usually the case along the Galactic plane, as the magnetic field vectors have orientation parallel to it. The free-free and the AME are known to be very lowly polarised (typically less than 1\%), therefore polarisation at this level of $\sim 0.35$~mK at 22.7~GHz could only arise from synchrotron or from thermal dust. The thermal dust template at this frequency derived from \commander shows polarised intensities below $0.05$~mK in this region. Therefore, we conclude that the bulk of this emission comes from synchrotron.

\begin{figure*}
\centering
\includegraphics[width=17cm]{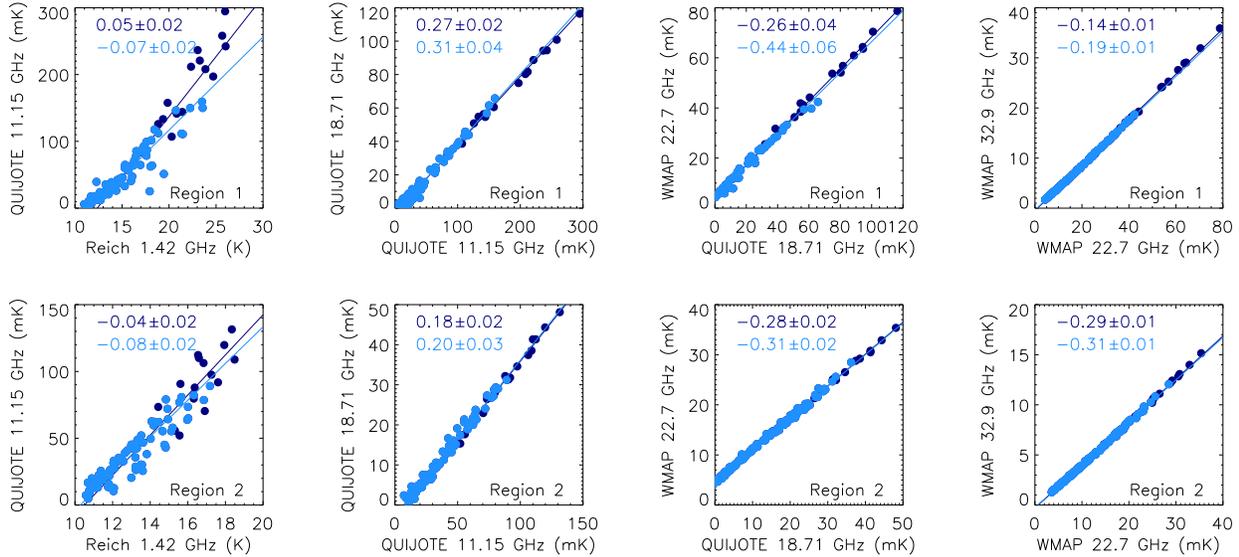}
\caption{\small Correlation plots in total intensity between different pairs of frequency bands, calculated in the two regions indicated in Fig.~\ref{fig:large_maps}: Region 1 (top panels) encompass an area of $32$~deg$^2$ along the Galactic plane to the west of W44, and Region 2 (bottom panels) subtends an equally large area but located to the east of W44. Dark-blue points correspond to individual pixels in these regions, while light-blue points correspond to the surviving pixels after masking out the two bright HII regions W43 and W47. From a fit to a linear polynomial we derive the spectral indices indicated in the top-left corner of each panel.}
\label{fig:tt}
\end{figure*}

The spectral properties of the diffuse resolved emission can be studied through correlation plots (see e.g. \citet{fuskeland14,cpp2015-25,irfan15}). We define the two boxes indicated in Fig.~\ref{fig:large_maps}, enclosing pixels with $|b|<2^\circ$, and with $25.7^\circ<l<33.7^\circ$ for `Region 1', which contains W43, and with $35.8^\circ<l<43.8^\circ$ for `Region 2', which contains W47. We have convolved all the frequency maps to a common angular resolution of $1^\circ$ and, in order to minimize the correlation between pixels, degraded them to $N_{\rm side}=128$ (pixel size of $0.46^\circ$). In Fig.~\ref{fig:tt} we show the resulting intensity-intensity correlation plots for different combinations of frequencies. Each dark-blue point represents one individual pixel, whereas the light-blue points represent the remaining pixels after excising those closer than $1^\circ$ to W43 or W47. By fitting the points to a linear polynomial, using the {\sc idl} routine {\sc mpfitexy} \citep{markwardt09}, which takes into account in the $\chi^2$ minimisation the errors in the two axes, we get the spectral indices indicated in the legend of each panel\footnote{Note that, as we are representing in these plots spectral intensities, the fit to a linear polynomial in this case gives the index of the spectrum of  spectral intensity, $\beta_{\rm T}={\rm log}(s)/{\rm log}(\nu_{\rm y}/\nu_{\rm x})$, where $s$ is the slope of the fit, and $\nu_{\rm x}$ and $\nu_{\rm y}$ are the frequencies of the data represented in the horizontal and vertical axes, respectively. As in the rest of the article we refer to the index of the flux density spectrum, in Fig.~\ref{fig:tt} we quote these quantities which, under the Rayleigh-Jeans approximation, are related with the previous ones through $\beta=\beta_{\rm T}+2$.}. In most cases the results are fully consistent with a free-free spectrum, which has spectral index of $-0.13$ at 10~GHz and $-0.15$ at 30~GHz. Note however the upturn of the spectrum between 11.1~GHz and 18.7~GHz, as expected in the case of the presence of AME, which is known to have a rising spectrum in this frequency range. On the other hand, in the two regions we see a steepening of the spectrum between 18.7~GHz and 22.7~GHz, which must occur if the AME spectrum peaks in the frequency range $\sim 19-20$~GHz.

\begin{figure*}
\centering
\includegraphics[width=17cm]{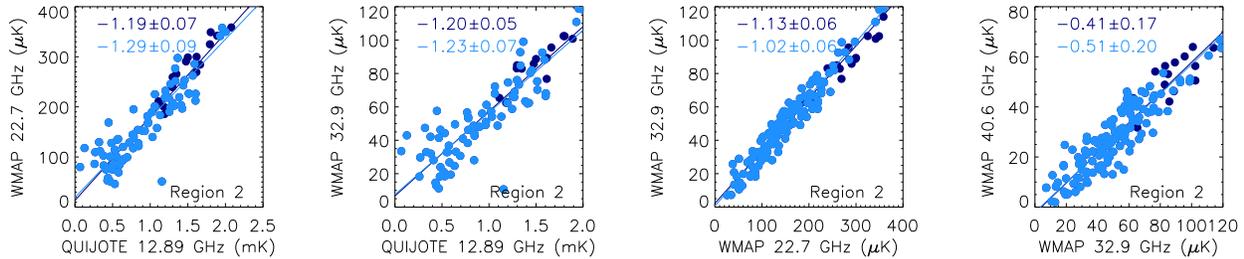}
\caption{\small Same as in Fig.~\ref{fig:tt} but for the polarised intensity in Region 2. }
\label{fig:pp}
\end{figure*}

In Fig.~\ref{fig:pp} we show correlation plots for the polarised intensity, after debiasing each individual pixel by applying the `modified asymptotic estimator' presented in \citet{plaszczynski14}. In this case we present only results for region 2. Region 1 has been avoided because of the presence of many pixels that are removed as a consequence of satellite contamination in the two lowest QUIJOTE frequency bands. Below 30~GHz the spectral indices are fully consistent with a synchrotron spectrum (typical synchrotron spectral indices are in the range $-0.6$ to $-1.6$ \citep{bennett13}). There is no polarised emission associated with W47, so masking this source does not result in a noticeable difference in the derived spectral indices. Contrary to what happens in intensity, in polarisation the spectrum do not flattens at frequencies between 12.9~GHz and 22.7~GHz, indicating that all the diffuse polarisation is due to synchrotron emission, with no hints of any AME polarisation. Although still compatible with synchrotron emission, the spectral index flattens at frequencies above 30~GHz. We interpret this as contamination from polarised thermal dust emission, which has a positive slope, and can start to be important at these frequencies.

\section{Total intensity emission from compact sources}\label{sec:intensity}

\subsection{Intensity flux densities}\label{subsec:intensity}
We calculate flux densities at each individual frequency through aperture photometry, a technique consisting in integrating all pixels inside a circle around the source, and subtracting a background level calculated as the median of all pixels enclosed in an external ring. This technique has been widely used to determine spectral energy distributions (SEDs) of AME sources \citep{lopez11,genova15}. As we will adopt the C-BASS flux at 4.76~GHz from \citet{irfan15}, we use exactly the same parameters of that paper: aperture radius of 60~arcmin, and background annulus defined by two concentric circles with radii 80 and 100~arcmin. For the central positions of the W43, W44 and W47 apertures we use their SIMBAD\footnote{{\tt http://simbad.u-strasbg.fr/simbad/}} coordinates, shown in Table~\ref{tab:sources}. The W43 and W44 coordinates differ from those of \citet{irfan15}, as they use $(30.8^\circ,-0.3^\circ)$,  $(34.7^\circ,-0.3^\circ)$ (Irfan, private communication). In W44, we scale the C-BASS flux density to the value associated with an aperture centred in our coordinates using the ratio between the HartRAO flux densities at 2.326~GHz (the nearest frequency to C-BASS) calculated using our and their coordinates. In W43 this scaling does not seem appropriate because the same ratio fluctuates above and below one in the \citet{haslam82}, \citet{berkhuijsen72}, \citet{reich86} and \citet{jonas98} maps. In this source, we keep the flux density quoted in \citet{irfan15}, and add a 10\% uncertainty to its error bar, which is of the order to the variation of flux densities in the low-frequency surveys when our and their coordinates are used. In W44 the scaling factor is $0.88$; the flux densities at 2.326~GHz of this source calculated using our and their coordinates are $253.9$~Jy and $287.2$~Jy, respectively. In the appendix~\ref{ap:correction_factors} we present a summary of these correction factors and others that have been mentioned before. 

\begin{table}
\begin{center}
\begin{tabular}{ccccc}
\hline\hline
\noalign{\smallskip}
Source & Type & $l$ (deg) & $b$ (deg) & $\theta$ (deg)\\
\noalign{\smallskip}
\hline
\noalign{\smallskip}
W43 & HII&$30.8$ & $-0.02$ & $1.8$ \\
\noalign{\smallskip}
W44 & SNR&$34.7$ & $-0.42$ & $0.5$ \\ 
\noalign{\smallskip}
W47 & HII&$37.8$ & $-0.19$ & $-$ \\
\noalign{\smallskip}
\hline\hline
\end{tabular}
\normalsize
\caption{Central coordinates (as taken from the SIMBAD database) and angular sizes of each of the three analysed sources.}
\label{tab:sources}
\end{center}
\end{table}

We show the C-BASS corrected flux densities in Table~\ref{tab:fluxes}, together with the values corresponding to the rest of the frequencies, which have been calculated directly on the maps at $1^\circ$ angular resolution, and at a pixelisation $N_{\rm side}=512$. The error bars listed in this table represent statistical uncertainties, and have been calculated through the RMS dispersion of the data in the background annulus. In appendices~\ref{ap:systematics} and \ref{ap:errors} we present a discussion of the possible impact of systematic uncertainties in our analyses. 

It must be noted that, owing to the coarse angular resolution, there might be significant contamination from the Galactic background emission. The subtraction of a median background level, as we do in our aperture photometry technique, may not be an optimal strategy to isolate the emission of the source, due to the strong gradient of the Galactic emission in the direction of the Galactic latitude. For this reason, in Table~\ref{tab:fluxes}, as well and in all the subsequent tables and figures, we have added the letter `r' to the name of the source, in order to emphasise that the derived flux densities correspond to regions that contain the sources W43, W44 and W47, and may actually result in an overestimation of the real flux densities of these sources due to background contamination. This may be particularly important for the SNR W44, which subtends an angle of $\approx 0.5^\circ$. On the other hand, W43 is a very large molecular cloud complex, subtending $\approx 1.3^\circ$ on the sky, so in this case the flux densities integrated in a radius of $1^\circ$ may represent a closer approximation to the true flux density of this source. In section~\ref{subsec:intensity_w44} we will try a different background subtraction in order to get a more reliable SED of the W44 SNR.

We have applied colour corrections for all surveys except the low-frequency ones ($0.408$ to $2.326$~GHz), where they can be safely neglected thanks to the narrow bandwidths of the detectors (typically $\Delta\nu/\nu<2\%$). As this correction obviously depends on the fitted model, we implemented an iterative process. In each iteration we integrate the fitted model on the QUIJOTE, \wmap and \planck bandpasses. We downloaded the \wmap and \planck bandpasses from the LAMBDA\footnote{{\tt http://lambda.gsfc.nasa.gov/product/map/dr5/bandpass\_get.cfm}} and the PLA\footnote{The LFI and HFI bandpasses are contained in the RIMO (Reduced Instrument Model) fits file, that can be found in the PLA.} archives, respectively. In the case of DIRBE, we used the colour correction tables given in the LAMBDA website\footnote{{\tt http://lambda.gsfc.nasa.gov/product/cobe/dirbe\_ancil\_cc\_get.cfm}}. Convergence is normally reached after the second iteration. The magnitude of the colour corrections is typically $\lesssim 0.5\%$ for QUIJOTE,  $\lesssim 1\%$ for \wmap and \planck-LFI, and $\lesssim 10\%$ for \planck-HFI and DIRBE. Our final fluxes, quoted in Table~\ref{tab:fluxes}, differ typically less than 10\% with those of \citet{irfan15}. These differences could be due to the different central coordinates, or to different colour correction strategies. However in W44, for frequencies larger than 143~GHz, our flux densities are systematically lower by around $50\%$.

\begin{table*}
\begin{center}
\begin{tabular}{cccccccc}
\hline\hline
\noalign{\smallskip}
Freq. &  \multicolumn{4}{c}{Flux density (Jy)}& Cal.  & Res.& Telescope/ \\
\noalign{\smallskip}
\cline{2-5}
\noalign{\smallskip}
(GHz) & W43r & W44r & W44 & W47r & (\%) & (arcmin) &  survey\\
\noalign{\smallskip}\hline\noalign{\smallskip}
 0.408   &    $ 503\pm  21$   &    $ 541\pm  23$   &   $ 379\pm 14$   &    $ 243\pm  22$  &  $10$  &51&     Haslam    \\
 0.82   &    $ 445\pm  18$   &    $ 382\pm  15$   &   $ 243\pm  9$   &    $ 213\pm  15$   &   $6$  &72&    Dwingeloo  \\
 1.42   &    $ 388\pm  17$   &    $ 329\pm  12$   &   $ 197\pm  7$   &    $ 197\pm  10$   & $5+5$  &34.2&      Reich    \\
 2.33   &    $ 461\pm  18$   &    $ 254\pm  16$   &   $ 179\pm 10$   &    $ 182\pm  14$   & $5+5$  &20&    HartRao    \\
 4.76   &    $ 400\pm  48$   &    $ 191\pm  15$   &   $   -      $   &    $ 166\pm  15$   & $5+5$  &43.8&   C-BASS    \\
11.15   &    $ 511\pm	9$   &    $ 192\pm   7$   &   $ 117\pm  7$   &    $ 174\pm   5$   &   $5$  &50.4&    QUIJOTE    \\
11.22   &    $ 510\pm	6$   &    $ 193\pm   6$   &   $ 118\pm  6$   &    $ 168\pm   5$   &   $5$  &53.2&    QUIJOTE    \\
12.84   &    $ 551\pm	7$   &    $ 204\pm   6$   &   $ 124\pm  6$   &    $ 178\pm   6$   &   $5$  &53.5&    QUIJOTE    \\
12.89   &    $ 544\pm  10$   &    $ 209\pm   7$   &   $ 124\pm  7$   &    $ 181\pm   6$   &   $5$  &50.8&    QUIJOTE    \\
16.75   &    $ 564\pm  10$   &    $ 212\pm   7$   &   $ 123\pm  7$   &    $ 187\pm   6$   &   $5$  &37.8&    QUIJOTE    \\
17.00   &    $ 546\pm	9$   &    $ 208\pm   6$   &   $ 125\pm  6$   &    $ 181\pm   6$   &   $5$  &39.1&    QUIJOTE    \\
18.71   &    $ 587\pm  11$   &    $ 221\pm   7$   &   $ 124\pm  7$   &    $ 193\pm   6$   &   $5$  &37.8&    QUIJOTE    \\
19.00   &    $ 576\pm  10$   &    $ 213\pm   7$   &   $ 128\pm  7$   &    $ 184\pm   6$   &   $5$  &39.1&    QUIJOTE    \\
 22.7   &    $ 548\pm  15$   &    $ 199\pm  10$   &   $ 116\pm  6$   &    $ 180\pm   9$   & $0.2$  &51.3&       \wmap    \\
 28.4   &    $ 542\pm  15$   &    $ 190\pm   9$   &   $ 102\pm  6$   &    $ 171\pm   8$   &$0.35$  &33.1&     \planck    \\
 32.9   &    $ 521\pm  14$   &    $ 177\pm   9$   &   $  99\pm  5$   &    $ 163\pm   8$   & $0.2$  &39.1&       \wmap    \\
 40.6   &    $ 480\pm  13$   &    $ 157\pm   8$   &   $  86\pm  5$   &    $ 148\pm   7$   & $0.2$  &30.8&       \wmap    \\
 44.1   &    $ 467\pm  12$   &    $ 152\pm   8$   &   $  84\pm  5$   &    $ 144\pm   6$   &$0.26$  &27.9&     \planck    \\
 60.5   &    $ 430\pm  11$   &    $ 141\pm   7$   &   $  78\pm  4$   &    $ 140\pm   6$   & $0.2$  &21.0&       \wmap    \\
 70.4   &    $ 449\pm  12$   &    $ 152\pm   8$   &   $  84\pm  5$   &    $ 152\pm   7$   &$0.20$  &13.1&     \planck    \\
 93.0   &    $ 560\pm  16$   &    $ 217\pm  12$   &   $ 124\pm  7$   &    $ 212\pm  12$   & $0.2$  &14.8&       \wmap    \\
  100   &    $ 620\pm  18$   &    $ 248\pm  14$   &   $ 146\pm  9$   &    $ 241\pm  14$   &$0.09~(+10)$  &9.7&     \planck    \\
  143   &    $1302\pm  44$   &    $ 606\pm  37$   &   $ 361\pm 23$   &    $ 572\pm  41$   &$0.07$  &7.3&     \planck    \\
  217   &    $4837\pm 172$   &    $2436\pm 149$   &   $1459\pm 93$   &    $2229\pm 165$   &$0.16~(+2)$  &5.0&     \planck   \\
  353   & $ (2.36\pm 0.08)\times 10^4$ & $ (1.19\pm 0.07)\times 10^4$ & $  7078\pm    452$             & $ (1.09\pm 0.08)\times 10^4$  &$0.78~(+5)$ &4.9&   \planck \\
  545   & $ (9.30\pm 0.33)\times 10^4$ & $ (4.49\pm 0.27)\times 10^4$ & $ (2.60\pm  0.17)\times 10^4$  & $ (4.16\pm 0.29)\times 10^4$  & $6.1$ &4.8&   \planck \\
  857   & $ (3.70\pm 0.13)\times 10^5$ & $ (1.64\pm 0.10)\times 10^5$ & $ (8.76\pm  0.65)\times 10^4$  & $ (1.58\pm 0.10)\times 10^5$  & $6.4$ &4.6& \planck \\
 1249   & $ (9.41\pm 0.31)\times 10^5$ & $ (3.76\pm 0.24)\times 10^5$ & $ (1.85\pm  0.15)\times 10^5$  & $ (3.84\pm 0.24)\times 10^5$  &$11.6$ &37.1&   DIRBE \\
 2141   & $ (1.88\pm 0.06)\times 10^6$ & $ (6.16\pm 0.42)\times 10^5$ & $ (2.67\pm  0.26)\times 10^5$  & $ (6.54\pm 0.36)\times 10^5$  &$10.6$ &38.0&  DIRBE \\
 2997   & $ (1.07\pm 0.03)\times 10^6$ & $ (3.00\pm 0.20)\times 10^5$ & $ (1.24\pm  0.13)\times 10^5$  & $ (3.20\pm 0.16)\times 10^5$  &$13.5$ &38.6&  DIRBE \\
 \noalign{\smallskip}
\hline\hline
\end{tabular}
\normalsize
\caption{Flux densities for the regions W43r, W44r and W47r. They have been calculated through aperture photometry in a ring of radius $60$~arcmin, and subtracting the median of the background calculated in a ring between 80 and 100~arcmin. Also shown are the flux densities of W44, for which we have subtracted a background level defined by two profiles calculated at two constant Galactic longitudes, as explained in section~\ref{subsec:intensity_w44}. The last three columns indicate the calibration uncertainties, angular resolution (FWHM) and the telescope or survey from which the data have been extracted. The C-BASS flux densities have been adopted from \citet{irfan15}.}
\label{tab:fluxes}
\end{center}
\end{table*}

\subsection{Characterisation of the free-free emission}
As it was mentioned in section~\ref{sec:ancillary_data}, in order to pin down the contribution from free-free emission to the low-frequency flux densities, we use two different templates. The first is a \planck map, derived from the \commander component separation technique \citep{cpp2015-10}, which contains in each sky pixel values of the electron temperature ($T_{\rm e}$) and of the emission measure ($EM$). We applied aperture photometry on these maps, using the same parameters that were used to obtain the flux densities. In Table~\ref{tab:ff} we show the average $T_{\rm e}$ in the aperture, and the derived values of $EM$ for each region. Using this information, we estimated the amplitude of the free-free emission at 1.4~GHz. The intervals indicated in Table~\ref{tab:ff} correspond to the $1\sigma$ confidence regions around the central value, where the error has been inferred from the standard deviation of the data in the background annulus.

\begin{figure}
\centering
\includegraphics[width=\columnwidth]{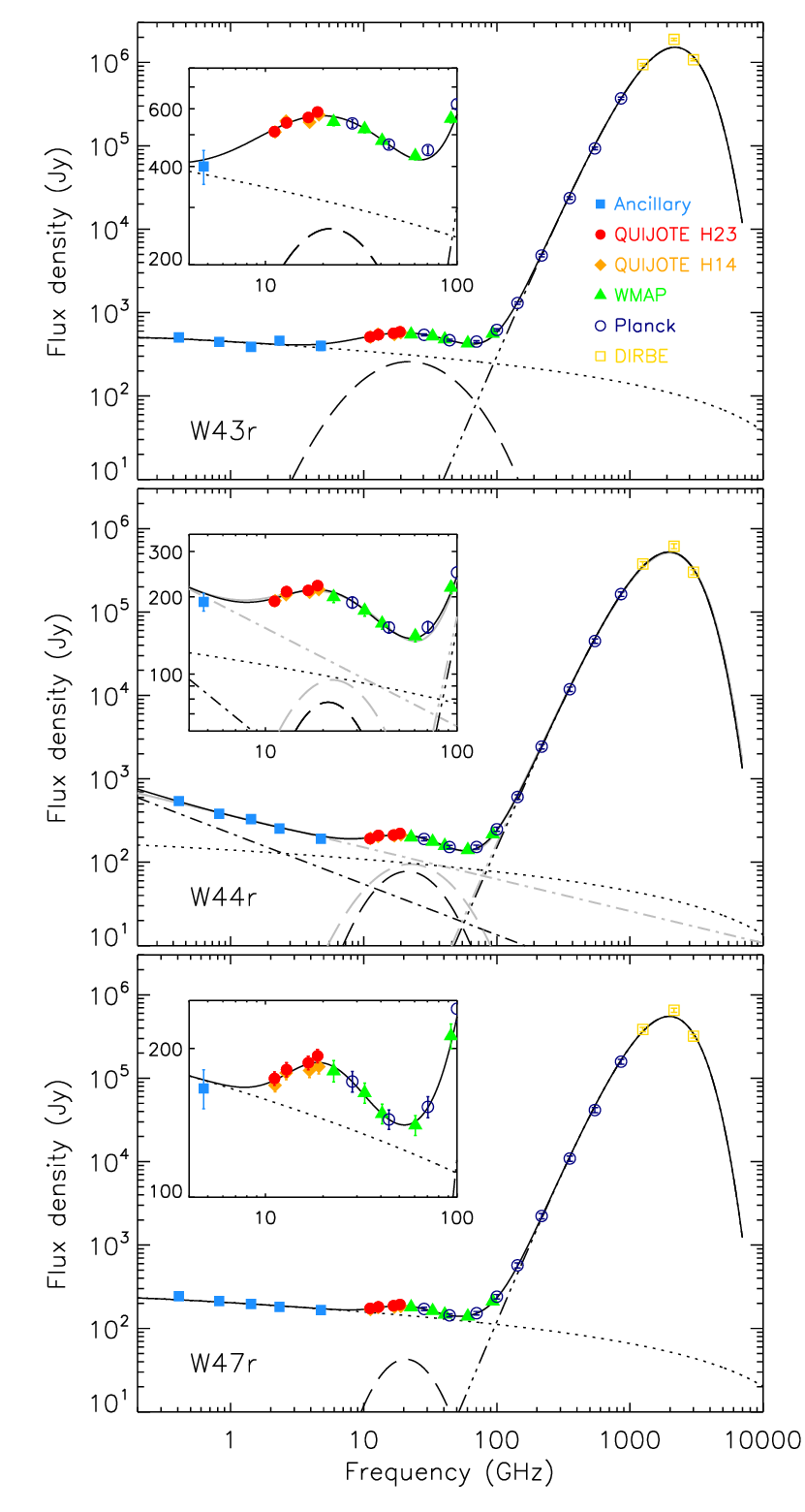}
\caption{\small Spectral energy distributions of the regions W43r (top), W44r (middle) and W47r (bottom). We represent eight QUIJOTE points coming from the four different horns, together with ancillary data including \wmap 9-year release, \planck second release and DIRBE. In all cases the excess of emission associated with AME clearly shows up at intermediate frequencies. A joint fit have been performed to all the data points, consisting of the following components: free-free (dotted line), synchrotron (only for the case of the region W44r; dotted-dashed line), spinning dust (dashed line) and thermal dust (dashed-triple-dotted line). The solid line represents the sum of all the components. The reduced chi-squared of the fits are $\chi^2_{\rm red}=5.4$, $1.0$ and $1.0$ respectively for W43r, W44r and W47r. In W44r we also tried a fit without free-free (grey lines), which results in $\chi^2_{\rm red}=1.5$.}
\label{fig:seds}
\end{figure}

The second template is a free-free map derived by \citet{alves12} using radio recombination lines from the HIPASS survey. They also delivered a map of the electron temperature. The free-free flux density and the electron temperatures for each region are shown in Table~\ref{tab:ff}. Here, the confidence interval corresponds to the quadratic sum of the dispersion of the data in the background annulus and a 15\% uncertainty that, following \citet{irfan15}, we assign to this free-free template. The $EM$ values have been derived from the average electron temperatures and from the free-free amplitudes.

\begin{table*}
\begin{center}
\begin{tabular}{ccccccccc}
\hline\hline
\noalign{\smallskip}
&& \multicolumn{3}{c}{\citet{cpp2015-10}} && \multicolumn{3}{c}{\citet{alves12}} \\
\noalign{\smallskip}
\cline{3-5}\cline{7-9}
\noalign{\smallskip}
Region && $\langle T_{\rm e}\rangle$ (K)  &  $EM$ (cm$^{-6}$~pc)  &  (S$_{1.4}$)$_{\rm ff}$ (Jy) && $\langle T_{\rm e}\rangle$ (K)  &  $EM$ (cm$^{-6}$~pc)  &  (S$_{1.4}$)$_{\rm ff}$ (Jy) \\
\noalign{\smallskip}\hline
W43r   && 6350 & 5888  & 794 -- 822  & &  6038  &  4020 -- 6190  &  446 -- 687 \\
\noalign{\smallskip}
W44r   &&  6208 &  1667 &  228 -- 244 & &  6636 &  990 --  1340 & 106 -- 144 \\
\noalign{\smallskip}
W47r   &&  6512  &1806 & 233 -- 245 & & 6757 & 1360 -- 1840 & 144 --195 \\
\hline\hline
\end{tabular}
\normalsize
\caption{Estimates of the flux density of the free-free emission at 1.4~GHz, derived from the \citet{cpp2015-10} map obtained through the \commander component-separation algorithm, and from the template derived from radio recombination lines observations by \citet{alves12}. These flux densities have been obtained by applying aperture photometry on these maps, using the same parameters as for the flux densities shown in Table~\ref{tab:fluxes}. In the first case we estimate the free-free flux density from the maps of electron temperature and emission measure. In the second case two maps are given, one for the electron temperature and another one for the flux density, and from the combination of the two we get an estimate of the emission measure. The allowed ranges account for the scatter of the data in the background annulus, and in the case of the \citet{alves12} map, for the 15\% uncertainty associated with the determination of the electron temperature.}
\label{tab:ff}
\end{center}
\end{table*}

\subsection{SED modelling}\label{subsec:intensity_w43_w44_w47}
The final flux densities corresponding to circular regions of radius $1^\circ$ centred in each source are represented versus the frequency in Fig.~\ref{fig:seds}. QUIJOTE flux densities derived from horns 2 and 3 are represented by red filled circles, whereas those derived from horns 1 and 4 are depicted by orange filled diamonds. Note the outstanding agreement between the same frequencies from different horns, a fact that confirms the reliability of our calibration strategy and of our map-making algorithm. It can be seen in Table~\ref{tab:fluxes} that this agreement is always within the error bars for W44r and W47r. In W43r this is also the case for all pairs of frequencies except for 16.7~GHz and 17.0~GHz, where the difference is $1.3\sigma$.

We perform a multi-component fit to all data points, consisting of four components: free-free and synchrotron emissions, which dominate the radio range, a model of spinning dust emission, that is important in the microwave range, and thermal dust emission, which clearly dominates the spectra in the far-infrared regime. We fix the shape of the free-free spectrum using the standard formulae given in \citet{per20}, adopting for the electron temperatures the values derived for each region from the \citet{alves12} map, which are indicated in Table~\ref{tab:ff}. The only free parameter associated with this component is therefore its amplitude, which is parameterised through the emission measure $EM$. The synchrotron spectrum is represented by its amplitude at 1~GHz, $S_{\rm sync}^{\rm 1 GHz}$, and its spectral index, $\beta_{\rm sync}$. This component is only considered in the case of the region W44r, whose emission is dominated by the SNR W44. The two other sources are HII regions and do not show synchrotron emission. In order to break possible degeneracies between the free-free and the synchrotron parameters in W44r, we set a flat prior on the emission measure $EM<1340$~cm$^{-6}$~pc, which comes from the upper bound of the $EM$ estimate from the \citet{alves12} map (see Table~\ref{tab:ff}). Following \citet{irfan15}, for the spinning dust emission we resort to the phenomenological model proposed by \citet{bonaldi07}, consisting of a parabola in the logarithmic space (log($S_{\nu})$--log$\nu$) described by three parameters: its slope at 60~GHz, $m_{60}$, which is associated with the width of the parabola, its central frequency, $\nu_{\rm AME}^{\rm peak}$, and its amplitude, $S_{\rm AME}^{\rm peak}$. Finally, the thermal dust is modelled as a single-component modified blackbody curve, $\tau_{250}(\nu/1200~{\rm GHz})^{\beta_{\rm d}} B_{\nu}(T_{\rm d})$, which depends on three parameters: the optical depth at $250~\mu$m, $\tau_{250}$, the emissivity spectral index, $\beta_{\rm d}$, and the dust temperature $T_{\rm d}$. Note that some authors (see e.g. Shetty et al. 2009) claim that an artificial anti-correlation between $\beta_{\rm d}$ and $T_{\rm d}$ arise when trying to fit SEDs to noisy data with limited frequency coverage. However, in our case we have high signal-to-noise data with typically 10 frequency points dominated by thermal dust emission, and therefore it is justified to simultaneously fit for the two parameters.

We then perform a joint fit of all these components to the observed data in which, for W43r and W47r, we jointly fit 7 parameters: $EM$, $m_{60}$, $\nu_{\rm AME}^{\rm peak}$, $S_{\rm AME}^{\rm peak}$, $\tau_{250}$, $\beta_{\rm d}$ and $T_{\rm d}$. In W44r we add the two synchrotron parameters, $S_{\rm sync}^{\rm 1 GHz}$ and $\beta_{\rm sync}$, so in this case we fit for 9 parameters. In order to explore the possibility of synchrotron emission associated with nearby SNRs, we have also tried to include this component in the fits of the W43r and W47r SEDs. However, we found that the chi-squared is not improved, so we conclude that the data do not favour the presence of synchrotron emission in either of these two regions.

The best-fit models, shown in Fig.~\ref{fig:seds}, provide an excellent description of all the observed data. The best-fit values for each parameter, together with their $1\sigma$ error bars, and reduced chi-squared, are quoted in Table~\ref{tab:parameters}. In order to account for the goodness of the fit, the error of each parameter has been multiplied by $\sqrt{\chi^2_{\rm red}}$. Note that the reduced chi-squared are very close to one for W44r ($\chi_{\rm red}^2=1.010$) and W47r ($\chi_{\rm red}^2=0.995$). In W43r ($\chi_{\rm red}^2=5.4$) the fit does not seem so good. The higher value of $\chi^2_{\rm red}$ here is driven by the thermal dust model. At frequencies $>70.4$~GHz the differences between the data points and the model are typically between $2\sigma$ and $6\sigma$. This could be indicative of more than one thermal dust component. In the calculation of the reduced chi-squared we used the error bars quoted in Table~\ref{tab:fluxes} which account for statistical uncertainties only. For comparison, we show inside brackets in Table~\ref{tab:parameters} the resulting reduced chi-squared when the systematic uncertainties specified in Table~\ref{tab:fluxes} are added in quadrature to the statistical uncertainties. In appendix~\ref{ap:systematics} we discuss these systematic uncertainties in detail. 

\subsubsection{Contribution of QUIJOTE data}

Following \citet{irfan15}, who compared their best fitted parameters before and after the introduction of the C-BASS data point, in Table~\ref{tab:parameters} we show an equivalent comparison with and without the inclusion of the eight QUIJOTE data points in the analysis. We may first compare our results without QUIJOTE with the results of \citet{irfan15} including C-BASS, as they are based on the same dataset (except for the fact that \citet{irfan15} exclude from the fit the 100~GHz and 217~GHz \planck data points due to being contaminated by CO emission, while we include them in the fit, after correcting for this emission). We always get lower AME peak frequencies, and a higher AME amplitude in W43r, but lower in W44r and W47r. This is probably due to the different levels of the best-fit free-free amplitudes. The widths of the spinning dust parabola (inversely proportional to $m_{60}$) are very similar in the three regions. In what concerns the thermal dust model, our values for $\beta_{\rm dust}$ are very similar, whereas for $T_{\rm dust}$ we get similar values in W43r and W47r but a lower value in the case of W44r. Finally, our values for $\tau_{250}$ are a factor 2 to 4 lower than those of \citet{irfan15}.  In W44r this can be explained by our lower values for the flux densities at frequencies above $\sim 143$~GHz.

The numbers in Table~\ref{tab:parameters} show that, as expected, the inclusion of the QUIJOTE data points affects mainly the spinning dust models. The thermal dust parameters barely change. The free-free $EM$ changes by less than 1\% in W43r and W47r. In W44r the inclusion of QUIJOTE data results in a $22\%$ increase of $EM$, while the synchrotron spectrum becomes steeper. On the other hand, the uncertainties on the AME peak frequencies and amplitudes are notably reduced thanks to QUIJOTE. Note for instance that the error bar of $\nu_{\rm AME}^{\rm peak}$ decreases from $6.7$ to only $1.0$~GHz in W44r. By including QUIJOTE data, the spinning dust parabola becomes narrower in W43r and in W44r, and wider in W47r. This is particularly important for W43r, where the fitted parabola looks much wider than the typical spinning dust spectra, as it was already pointed out by \citet{irfan15}. The peak frequency increases in W43r but decreases in W44r and in W47r. Finally, the QUIJOTE data makes the AME amplitude lower in W43r but higher in W44r and W47r.

\subsubsection{Free-free and synchrotron emissions}

The free-free spectrum provides an excellent fit to the W43r and W47r radio data. The best-fit $EM$ for W43r is slightly below the expected range derived from the \citet{alves12} maps. In W47r the best-fit $EM$ is slightly above the expected range, but still compatible given the error bar. In W44r we put the constraint $EM<1340$~cm$^{-6}$~pc, but the best-fit value, $EM=1264\pm 22$~cm$^{-6}$~pc, lies inside the allowed interval derived from the \citet{alves12} template. Our fitted synchrotron spectral index $\beta_{\rm sync}=-0.61\pm 0.04$, agrees with that of \citet{irfan15}, $\beta_{\rm sync}=-0.57\pm 0.08$, but is smaller than the value of $\beta_{\rm sync}=-0.37$ obtained by C07 and by \citet{green14}. It must be noted however that these fits are based on data at much finer angular resolutions, and do not consider a free-free component. As it was pointed out above, at an angular scale of $1^\circ$ we could have a contribution to the measured flux densites from nearby sources or from the background. In section~\ref{subsec:intensity_w44} we will try to carefully take into account these effects. 

If we do not include the free-free component in the fit, we get $S_{\rm sync}^{\rm 1 GHz}=364\pm 9$~Jy and $\beta_{\rm sync}=-0.38\pm 0.02$, now fully compatible with the value of $\beta_{\rm sync}=-0.37\pm 0.02$ from C07. In this case, the AME becomes stronger, $S_{\rm AME}^{\rm peak}= 94.9\pm 9.2$~Jy, at the cost of a poorer reduced chi-squared of $\chi^2_{\rm red}=1.51$, as compared to $\chi^2_{\rm red}=1.01$ when the free-free is introduced in the fit. In order to assess if the data favours the presence of free-free we can use the Bayesian information criterion \citep{schwarz78} BIC$=\chi^2+k{\rm log}N$, where $k$ is the number of parameters and $N$ the number of data points. When the free-free component is brought into the model we get an improvement of $\Delta$BIC$=-8.1$, which means strong evidence in favour of the free-free component. Note also that the RRL map of \citet{alves12} shows emission at the position of W44, which is further evidence of the presence of free-free.

\subsubsection{AME}

The data represented in Fig.~\ref{fig:seds} show very clearly the presence of AME, first discovered by \citet{irfan15}. At $18.7$~GHz AME is detected respectively at $21.2\sigma$, $10.2\sigma$ and $7.7\sigma$ in W43r, W44r and W47r. The eight QUIJOTE points confirm the downturn of the AME spectrum at low frequencies. As it was already discussed in section~\ref{subsec:intensity}, we emphasise that an important fraction of the AME could actually come from background Galactic emission rather than from the sources themselves. In section~\ref{subsec:intensity_w44} we will try to quantify more accurately the real AME associated with the SNR W44.

Contrary to other regions, like G159.6-18.5 in the Perseus molecular complex where the AME dominates the SED at frequencies $10-50$~GHz, in W43r, W44r and W47r the emission is always dominated by the free-free in this frequency range. In W43r and W44r the AME maximum flux densities are close to the free-free emission at 22.7~GHz, but still slightly below. In any case, the AME flux densities are rather high, with peak values of $\approx 280$~Jy in W43r and $\approx 80$~Jy in W44r, much higher than the maximum of $\approx 35$~Jy which is found in G159.6-18.5 \citep{genova15}. We obtain AME residual flux densities at 22.7~GHz (derived by subtracting from the measured \wmap flux densities at 22.7~GHz the rest of the components resulting from our fitted model evaluated at the same frequency) of $238\pm 16$, $68\pm 11$ and $38\pm 9$~Jy, respectively for W43r, W44r and W47r. These values are considerably higher than those extracted from the \commander AME template at 22.8~GHz through equivalent aperture photometry integrations: $65\pm 5$, $35\pm 5$ and $29\pm 5$~Jy, respectively. On the other hand, the values of Table~\ref{tab:ff} show that the free-free amplitudes predicted by \commander are higher than both the predictions from the RRL survey of \citet{alves12} and than the values derived form our models. It therefore seems clear that \commander overestimates the free-free emission and underestimates the AME associated with these regions. This highlights the important role of the QUIJOTE data between 10 and 20~GHz, tracing the downturn of the AME spectrum at frequencies below $\approx 20$~GHz, and in turn crucially helping to better determining the real AME amplitude.

It is common practice in the literature to parameterise the AME amplitude as the ratio between the AME peak and the $100~\mu$m (2997~GHz) flux densities, which should be proportional to the dust column density along the line of sight \citep{pip15}. We find these ratios, usually referred to as AME emissivities, to be $(2.41\pm 0.07)\times 10^{-4}$, $(2.59\pm 0.25)\times 10^{-4}$ and $(1.34\pm 0.10)\times 10^{-4}$, respectively for W43r, W44r and W47r. The value for W47r is compatible with the weighted average of  $(1.10\pm 0.21)\times 10^{-4}$ found by \citet{todorovic10} in a sample of nine HII regions. \citet{pip15} found a weighted average over 98 AME sources of $(1.47\pm 0.11)\times 10^{-4}$. However, they argue that this value could be biased by some sources with relatively small error bars and low emissivities, so they also quote the unweighted average, which is $(5.4\pm 0.6)\times 10^{-4}$. This is higher than what we find in our three regions, as it is also higher the emissivity of $6.2\times 10^{-4}$ found by \citet{davies06} in diffuse AME regions at high Galactic latitudes.

\begin{figure*}
\centering
\includegraphics[width=16cm]{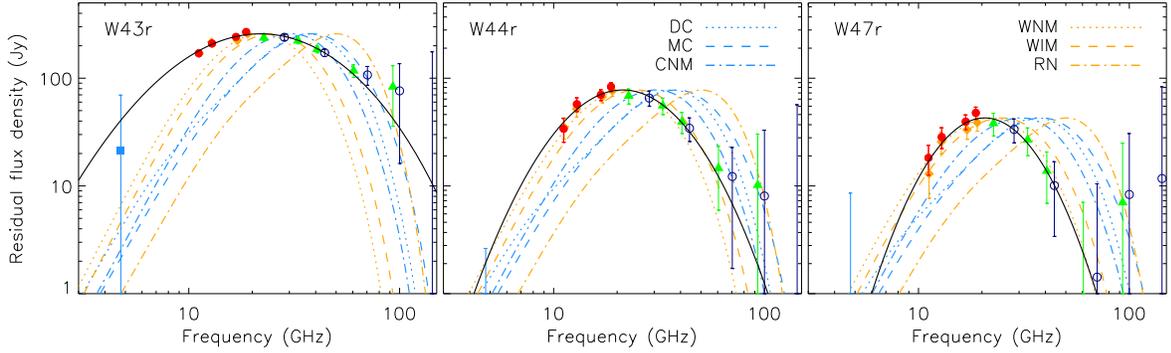}
\caption{\small Residual AME spectra of W43r (left), W44r (middle) and W47r (right), after subtracting the best-fit models for the free-free, synchrotron (only for W44) and thermal dust components shown in Fig.~\ref{fig:seds}. The colour coding for the observed flux densities is the same as in Fig.~\ref{fig:seds}. The solid line represents the best-fit AME model of \citet{bonaldi07}. The colour curves in the three panels represent the spinning dust models of \citet{draine98} for six different physical environments (indicated in the legends of the centre and right pannels), normalised to the same amplitude as the solid curve.}
\label{fig:res_seds}
\end{figure*}

The inset plots of Fig.~\ref{fig:seds}, and also the plots of Fig.~\ref{fig:res_seds} showing the residual AME flux densities and the best-fit spinning dust models, demonstrate that we manage to get very accurate fits of the spinning dust spectra. However, we have to bear in mind that we are using an AME model that is not physically motivated. In order to study the physical reliability of the fitted models, in Fig.~\ref{fig:res_seds} we plot our best-fit spectra together with the spinning dust models of \citet{draine98} for different environments, normalised to the same amplitudes. This confirms that the parabola for W43r is too wide. On the other hand, it seems that for W44r and W47r, the warm ionised medium (WIM) and the warm neutral medium (WNM) provide, respectively, the most similar spectrum to the fitted parabola. We tried to use the WIM model in the fit of the W44r data and got a worse fit with $\chi^2_{\rm red}=1.57$. The synchrotron spectrum becomes in this case flatter, with $\beta_{\rm sync}=-0.49$, in an attempt to better fit the QUIJOTE points at 11 and 13~GHz, whose flux densities are too high as to be reproduced by the WIM spinning dust model. As a result, the 2.33~GHz and the 4.76~GHz data points are notably below the best-fit model. A more exhaustive study in this direction would require the use of the {\sc spdust} code\footnote{{\tt http://www.tapir.caltech.edu/$\sim$yacine/spdust/spdust.html}} \citep{ali09,silsbee11} to try tweaking some of the physical parameters of the WIM environment, with the aim to find a spinning dust model that reproduces better the observed data. This study could be the goal of future work.

\begin{table*}
\begin{center}
\begin{tabular}{lcccccccc}
\hline\hline
\noalign{\smallskip}
  & \multicolumn{2}{c}{W43r} && \multicolumn{2}{c}{W44r}  && \multicolumn{2}{c}{W47r} \\
\noalign{\smallskip}
\cline{2-3}\cline{5-6}\cline{8-9}
\noalign{\smallskip}
 Parameter & With QUIJOTE & Without && With QUIJOTE & Without && With QUIJOTE & Without \\
  \noalign{\smallskip}\hline\noalign{\smallskip}
                $EM$  (cm$^{-6}$~pc) &   3911  $\pm$    68  &   3882  $\pm$   126  &&   1264  $\pm$    22  &    999  $\pm$    42  &&   1849  $\pm$    20  &   1854  $\pm$    37   \\
  \noalign{\smallskip}
  $S_{\rm sync}^{\rm 1 GHz}$ (Jy) &                 $-$  &                 $-$  &&    222  $\pm$     7  &    255  $\pm$     9  &&		  $-$  &		 $-$   \\
  \noalign{\smallskip}
              $\beta_{\rm sync}$  &                 $-$  &                 $-$  &&  -0.61  $\pm$  0.04  &  -0.52  $\pm$  0.04  &&		  $-$  &		 $-$   \\
  \noalign{\smallskip}
                        $m_{60}$  &   1.56  $\pm$  0.36  &   1.18  $\pm$  0.51  &&   3.38  $\pm$  1.36  &   2.64  $\pm$  3.85  &&   5.21  $\pm$  1.41  &   5.75  $\pm$  1.16   \\
  \noalign{\smallskip}
$\nu_{\rm AME}^{\rm peak}$ (GHz)  &   22.2  $\pm$   1.1  &   20.5  $\pm$   6.0  &&   21.4  $\pm$   1.0  &   23.7  $\pm$   6.7  &&   20.7  $\pm$   0.9  &   22.6  $\pm$   2.7   \\
  \noalign{\smallskip}
   $S_{\rm AME}^{\rm peak}$ (Jy)  &  258.1  $\pm$   6.9  &  260.3  $\pm$  15.9  &&   77.7  $\pm$   5.5  &   73.3  $\pm$   9.2  &&   42.8  $\pm$   2.3  &   39.7  $\pm$   5.6   \\
  \noalign{\smallskip}
               $\beta_{\rm dust}$ &   1.75  $\pm$  0.05  &   1.78  $\pm$  0.06  &&   1.75  $\pm$  0.04  &   1.74  $\pm$  0.05  &&   1.87  $\pm$  0.04  &   1.87  $\pm$  0.05   \\
  \noalign{\smallskip}
              $T_{\rm dust}$ (K)  &   22.2  $\pm$   0.7  &   22.0  $\pm$   0.7  &&   20.1  $\pm$   0.4  &   20.2  $\pm$   0.5  &&   19.6  $\pm$   0.4  &   19.6  $\pm$   0.5   \\
  \noalign{\smallskip}
  $\tau_{250}$ ($\times 10^{-3}$) &   4.02  $\pm$  0.50  &   4.18  $\pm$  0.55  &&   2.25  $\pm$  0.21  &   2.21  $\pm$  0.24  &&   2.49  $\pm$  0.23  &   2.49  $\pm$  0.26   \\
  \noalign{\smallskip}
              $\chi^2_{\rm red}$  &   5.4  (0.9)         &   6.5  (1.3)         &&  1.0 (0.3)	        &  1.4 (0.4)           &&  1.0 (0.3)           &  1.3 (0.4)      \\              
\noalign{\smallskip}
\hline\hline
\end{tabular}
\normalsize
\caption{Best-fit model parameters for the three regions. We compare the two cases in which we include and exclude the QUIJOTE flux densities from the fit. The emission measure $EM$ defines the amplitude of the free-free emission. The synchrotron spectrum is defined its amplitude at 1~GHz, $S_{\rm sync}$, and its spectral index $\beta_{\rm sync}$. This component is fitted only to the W44 SED, as the other two objects are HII regions whose emissions are dominated by free-free at low frequencies. The AME component is fitted using the phenomenological model of \citet{bonaldi07} consisting in a parabola in the log$S$-log$\nu$ plane which depends on three parameters: its slope at 60~GHz $m_{60}$, the peak AME flux $S_{\rm AME}^{\rm peak}$, and the frequency corresponding to this flux $\nu_{\rm AME}^{\rm peak}$. The thermal dust spectrum is represented by a modified-black body law, characterised by three free parameters: the optical depth at 250~$\mu$m ($\tau_{250}$), the emissivity spectral index ($\beta_{\rm dust}$) and the dust temperature ($T_{\rm dust}$). In the last line we show the reduced chi-squared of each fit and, in parentheses, the values of the best-fit models when the calibration errors are added in quadrature to the statistical errors.}
\label{tab:parameters}
\end{center}
\end{table*}

\subsection{Re-evaluation of the intensity SED of W44}\label{subsec:intensity_w44}
As it was discussed in section~\ref{subsec:intensity}, the large aperture size together with the coarse angular resolution of the data (FWHM$=1^\circ$) could result in significant contamination from nearby sources or from diffuse Galactic emission. While not so important in W47, which is $\approx 1.3^\circ$ across, this may be particularly relevant for W44 given its angular size of $\approx 0.5^\circ$. In fact, our W44r flux densities below 10~GHz listed in Table~\ref{tab:fluxes} are systematically larger than those of C07, which may correspond to a more accurate estimate of the true flux density of the source as they come from high-angular resolution observations. At 1.4~GHz they give values in the range $180-270$~Jy, whereas at the same frequency we obtained $329\pm 12$~Jy. 

In section~\ref{subsec:intensity} we added the `r' label to the names of the three regions, W43r, W44r and W47r, precisely to highlight that their measured flux densities could have significant contributions from the diffuse background or from compact objects different from the W44 SNR or from the W43 and W47 HII regions themselves. In order to try to better isolate the emission from the SNR W44, to which we will refer to simply as W44, we now attempt a more refined subtraction of the background. In sources on the Galactic plane, it is evident that the median level calculated in the external ring may not be representative of the real average background in the aperture (see the intensity maps in Fig.~\ref{fig:iqu_maps}). To try to estimate the background contribution to each pixel in a more reliable way, we define two profiles of the Galactic emission as a function of $b$ through two cuts of the Galactic plane at constant $l$. These two cuts are symmetrically located around the source at $l_{\rm c1}=l_{\rm W44}-\Delta l$ and at $l_{\rm c2}=l_{\rm W44}+\Delta l$, where $l_{\rm W44}=34.7^\circ$. We then assume that at each individual frequency the background around the source is a function of $b$ only, and is given by the average of the two profiles given by the previous cuts. After trying different values for $\Delta l$ we concluded that $\Delta l=1.5^\circ$ rendered low-frequency flux densities compatible with those of C07. Also, the maps in Fig.~\ref{fig:iqu_maps} show that $l_{\rm c1}=33.2^\circ$ and $l_{\rm c2}=36.2^\circ$ are well away from W44 or any other point sources, and may define profiles that should be representative of the real diffuse Galactic background. 

\begin{figure}
\centering
\includegraphics[width=\columnwidth]{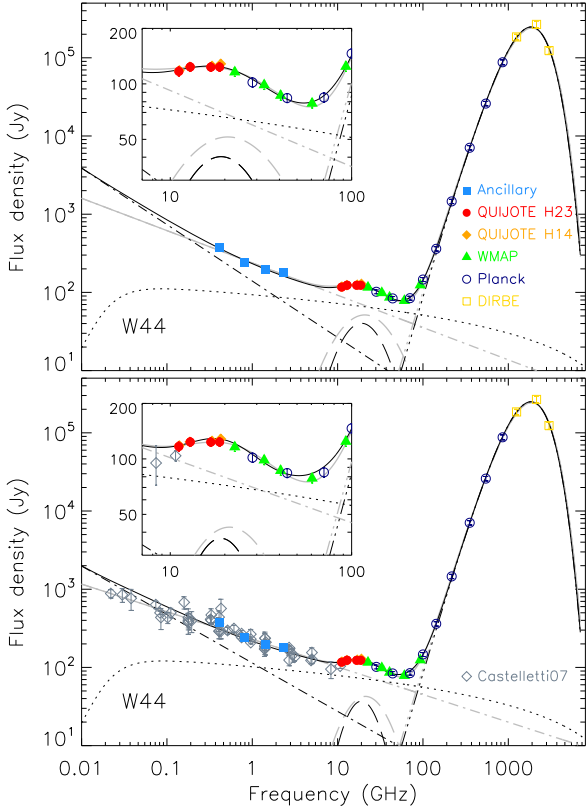}
\caption{\small Spectral energy distribution of the SNR W44. At each frequency we performed a more careful subtraction of the background level than in Fig.~\ref{fig:res_seds}, in order to try to better isolate the emission from the SNR (see text for details). We performed a joint fit to the observed data, including free-free (dotted line), synchrotron (dotted-dashed line), spinning dust (dashed line) and thermal dust (dashed-triple-dotted line) emissions. The solid line represents the sum of all the components. We have performed a fit with (black lines) and without (grey lines) a free-free component. In the upper panel we considered only our inferred flux densities, whereas in the lower panel we combined our measurements with those inferred from high-angular resolution observations (C07). In all these fits, the AME is still present, although at a lower level than in Fig.~\ref{fig:res_seds}.}
\label{fig:sed_w44}
\end{figure}

\subsubsection{Free-free and synchrotron emissions}

The final flux densities derived using this improved background subtraction are listed in Table~\ref{tab:fluxes} and represented in Fig.~\ref{fig:sed_w44}. They exhibit a good agreement with the compilation of public data of C07. Note that the QUIJOTE flux density at 11~GHz shows not only consistency between the two horns, but also (at $1.4\sigma$) with the measurement at 10.4~GHz of $104\pm 7$~Jy, which comes from \citet{kundu72}. Although the low-frequency emission of the W44 SNR should in principle be dominated by synchrotron, following what we already did in section~\ref{subsec:intensity_w43_w44_w47} we perform a fit with and without free-free emission, plus synchrotron, AME and thermal dust. In the upper and lower panels of Fig.~\ref{fig:sed_w44} we show the resulting fits to our inferred data, and to the combination of our and C07 data points, respectively. There are some clear outliers in the 55 data points of C07. In order to avoid them we first remove 4 points with very low errors that are too far from the fitted model and render the $\chi^2_{\rm red}$ too high, and later apply a clipping that removes two more points that lie more than $3\sigma$ from the initially fitted model. In the lower panel of Fig.~\ref{fig:sed_w44} we only plot the 49 remaining points that we use in the fit. 

The best-fit parameters are listed in Table~\ref{tab:parameters_w44}. When we use only our data, we get now a much lower synchrotron amplitude ($S_{\rm sync}^{\rm 1 GHz}=140\pm 4$ and $238\pm 5$~Jy, respectively with and without free-free component included in the fit) than for W44r ($S_{\rm sync}^{\rm 1 GHz}=222\pm 7$ and $364\pm 9$~Jy), indicating that an important fraction of the low-frequency emission measured for W44r may be associated with the diffuse Galactic emission rather than with the source itself. The spectral index remains almost unchanged: $\beta_{\rm sync}=-0.41\pm 0.02$ now for W44, compared to $-0.38\pm 0.02$ for W44r without free-free. Table~\ref{tab:parameters_w44} also shows that the inclusion of C07 data points do not affect the fitted parameters significantly. The synchrotron spectrum becomes now slightly flatter, $\beta_{\rm sync}=-0.35\pm 0.01$, and as a result the AME amplitude decreases.

This latter spectral index is consistent with previous studies \citep{castelletti07,green14,onic15,pip31}. In \citet{pip31} they see an excess of emission at 28.4~GHz and a deficit at 70.4~GHz with respect to the synchrotron law extrapolated from low frequency. We believe that the 28.4~GHz flux density could be affected by AME. However, they argue that this point could be contaminated by diffuse Galactic emission, and explain the lower flux density at 70.4~GHz through a possible steepening of the synchrotron spectrum. In that article, and also in \citet{onic15}, they fit a combination of flux densities at different frequencies that have been obtained from data at different angular resolutions and using different techniques. This may not be adequate, as there could be completely different background contributions at different frequencies. Also, in the aperture photometry applied in \citet{pip31} they use at each frequency an aperture radius that is proportional to the convolution between the source and the beam. We agree that through this technique the source will contribute equally to the flux densities at different frequencies. However, in cases of bright background emission, as is this case, the amount of flux coming from the surrounding background that leaks into the aperture will change at each frequency. For this reason, we think a better strategy is to degrade all the maps to the same angular resolution, and use the same aperture size for all frequencies, as we do here.

For consistency with the analyses presented in section~\ref{subsec:intensity_w43_w44_w47}, we also try to fit to the W44 data a model including a free-free component. The fitted models are depicted in Fig.~\ref{fig:sed_w44} by black lines, in contrast to the models without free-free component which are represented by grey lines. The resulting best-fit parameters are quoted in Table~\ref{tab:parameters_w44}, and show that, as expected, the synchrotron spectrum steepens while the AME amplitude gets slightly reduced. We get a notably lower reduced chi-squared of $\chi^2_{\rm red}=0.55$. The application of the same Bayesian model selection criterion as in section~\ref{subsec:intensity_w43_w44_w47} leads to $\Delta$BIC$=-7.2$, meaning strong evidence for the presence of free-free emission. It is well known that SNRs have radio spectra dominated by synchrotron emission. However there are several scenarios that could explain the presence of free-free emission associated with these objects. \citet{onic12} argue that there could be ``radio thermally active'' SNRs that have expanded into a high-density ISM, e.g. molecular cloud environment, and could host detectable thermal bremsstrahlung emission. This is more-likely to occur in evolved SNRs with mixed morphology. W44 is not very old ($\sim 20,000$ years) but has in fact a mixed morphology, characterised by shell-like structure in radio and bright interiors due to thermal X-ray emission. In the case of W44, \citet{seta98} have found six giant molecular clouds surrounding this SNR, some of which seem physically interacting with it. Furthermore, C07 identified a strong flattening of the spectral index between 0.074 and 0.324~GHz in a region coincident with the very nearby HII region G034.7-00.6  \citep{paladini03}\footnote{G034.7-00.6 is the closest counterpart in the \citet{paladini03} catalogue to the region where C07 see the flattening of the spetral index. However, note that there is also another HII region in that catalogue, G034.7-00.5, with a higher flux density at 2.7~GHz (172.0~Jy, compared to 21.3~Jy for G034.7-00.6).}. They also found that this HII region is limited to the east by an annular photo dissociation region (PDR) whose 8$\mu$m emission is dominated by polycyclic aromatic hydrocarbons (PAHs). This region could be generating free-free emission, and in turn could also contribute to the observed AME. C07 also derive a map of the spectral index between $0.3$~GHz and $1.4$~GHz, which shows it to be between $-0.4$ and $-0.5$ in the periphery of the SNR, but considerably flatter in most of the diffuse interior, with some filaments reaching $\beta\sim 0.1$. This could be indicative of thermal emission from the interior of the SNR, which would show up more clearly in our SED at frequencies $\gtrsim 10$~GHz, a range not covered by the C07 analysis.

However, even if the reduced chi-squared clearly decreases and the BIC test strongly supports the presence of free-free emission, this component is not confirmed by the very low-frequency data of C07. Although these data, below $0.4$~GHz, have not been scaled to the same \citet{baars77} flux scale as the other data points, they do not exhibit a large scatter, so we yet decided to bring them into the fit. In this case, the inclusion of the free-free do not improve the chi-squared, meaning that this component is not favoured by the data. However, as we will see in section~\ref{subsec:sync_pol_w44}, the polarisation SED of W44 which, thanks to the lack of other components, is a much neater representation of the synchrotron spectrum, gives $\beta_{\rm sync}=-0.62\pm 0.03$ (see Table~\ref{tab:parameters_w44_pol}). This value goes in favour of the presence of a free-free contribution to the intensity SED, as in this case we get $\beta_{\rm sync}=-0.72\pm 0.04$ when we perform a fit to our inferred data, a value that is at $2.0\sigma$ with respect to the polarisation SED. We therefore perform a fit to the data including C07 fixing the spectral index at $\beta_{\rm sync}=-0.62$. Although with a higher reduced chi-squared, $\chi^2_{\rm red}=2.06$, the fitted model provides a good description of the data, as shown in the lower panel of Fig.~\ref{fig:sed_w44}. It must also be noted that here we are combining data coming from different observing techniques and angular resolutions, and as a consequence the chi-squared may not be a highly reliable indicator of what is the model best describing these data. 

Another argument that supports the contribution of free-free emission to the SED comes from the study of the W44 $\gamma$-ray spectrum performed by \citet{cardillo14}. The model that best describes their data is a broken power-law proton energy distribution, with spectral index $p_1=2.2\pm 0.1$ at low energies and $p_2=3.2\pm 0.1$ at high energies. This corresponds to synchrotron spectral indices in intensity of $\beta_1=-0.60\pm 0.05$ and $\beta_2=-1.10\pm 0.05$, respectively. They tried to fit to their data leptonic emission with $p=1.74$, as derived from the value $\beta=-0.37$ of C07, but this spectral index could not fit their low-energy data in any way. Also the hard synchrotron spectrum of C07 imposes some difficulties in the models proposed by \citet{cardillo16} based in the re-acceleration and compression of CRs. Note that the low-energy spectral index $\beta_1$ of \citet{cardillo14} is consistent with the value derived from the fit to our intensity SED with a free-free component, $\beta_{\rm sync}=-0.72\pm 0.04$, and with the polarisation SED, $\beta_{\rm sync}=-0.62\pm 0.03$.  However proton and electron spectra do no have necessarily the same spectral index. In any case, the steeper synchrotron spectral index that we derive here may have important implications on the models discussed in \citet{cardillo14} and \citet{cardillo16}.

\begin{table*}
\begin{center}
\begin{tabular}{lcccccc}
\hline\hline
\noalign{\smallskip}
  && \multicolumn{5}{c}{W44} \\
\noalign{\smallskip}
\cline{3-7}
\noalign{\smallskip}
&& \multicolumn{2}{c}{Excluding C07}  &&  \multicolumn{2}{c}{Including C07}  \\
\noalign{\smallskip}
\cline{3-4}\cline{6-7}
\noalign{\smallskip}
Parameter  &&  With free-free & Without free-free && With free-free & Without free-free  \\
\noalign{\smallskip}
\hline
\noalign{\smallskip}
$EM$  (cm$^{-6}$~pc) && $  842\pm 13  $    &       $-$	        && $  919\pm 22  $   &	      $-$   \\
\noalign{\smallskip}
      $S_{\rm sync}^{\rm 1 GHz}$  (Jy)        && $  140\pm 4   $    &  $  238\pm 5$      && $  114\pm 4	$   &	$229\pm 3$  \\
\noalign{\smallskip}
              $\beta_{\rm sync}$              && $-0.72\pm 0.04$    &  $-0.41\pm 0.02$   && $-0.62$ (fixed)         &	$-0.35\pm 0.01$\\
\noalign{\smallskip}
                        $m_{60}$              && $ 4.28\pm 1.17$    &  $ 2.25\pm 0.16$   && $ 7.29\pm 2.37$   &	$3.62\pm 0.87$   \\
\noalign{\smallskip}
$\nu_{\rm AME}^{\rm peak}$ (GHz)              && $ 19.1\pm  0.7$    &  $ 21.0\pm 0.9 $   && $ 19.1\pm  1.1$   &	$21.4\pm 1.0$	\\
\noalign{\smallskip}
   $S_{\rm AME}^{\rm peak}$ (Jy)              && $ 40.3\pm  2.7$    &  $ 51.5\pm 5.7 $   && $ 37.1\pm  3.5$   &	$42.6\pm 4.0$	\\
\noalign{\smallskip}
              $\beta_{\rm dust}$              && $ 1.76\pm 0.03$    &  $ 1.65\pm 0.05$   && $ 1.79\pm 0.07$   &	$1.69\pm 0.05$   \\
\noalign{\smallskip}
                  $T_{\rm dust}$ (K)          && $ 18.8\pm 0.4 $    &  $ 19.7\pm 0.6 $   && $ 18.7\pm 0.7 $   &	$19.4\pm 0.6$	\\
\noalign{\smallskip}
  $\tau_{250}$ ($\times 10^{-3}$)             && $ 1.43\pm 0.12$    &  $ 1.19\pm 0.14$   && $ 1.49\pm 0.23$   &	$1.27\pm 0.15$   \\
\noalign{\smallskip}
              $\chi^2_{\rm red}$              &&      $0.55$        &         $1.05$     &&     $2.06$	    &	$1.23$   \\
\noalign{\smallskip}
\hline\hline                           
\end{tabular}
\normalsize
\caption{Best-fit parameters resulting from the fit of the model to the SNR W44 flux densities derived in section~\ref{subsec:intensity_w44}, and plotted in Fig.~\ref{fig:sed_w44}. We independently consider a model with and without free-free emission, that we fit to our derived flux densities, and to the combination of these values with the flux densities of C07. The resulting models are plotted in Fig.~\ref{fig:sed_w44}. In the last line we show the reduced chi-squared of each fit.}
\label{tab:parameters_w44}
\end{center}
\end{table*}

\subsubsection{AME}

Regardless the combination of components that is used to describe the radio data, in this improved analysis the AME is still clearly detected in the microwave range, although with an amplitude a factor $\approx 40\%$ lower than for W44r. We attempted a fit to the combination of our and C07 data without AME, and fixing $\beta_{\rm sync}=-0.62$, and got a much worse fit with $\chi^2_{\rm red}=3.1$. The introduction of AME leads to $\Delta$BIC$=-65$, which means very strong evidence of this component. Therefore, even if there seems to be some diffuse Galactic AME (see section~\ref{sec:diffuse}), it follows from this analysis that there is AME intrinsic to W44. If confirmed, this would be the first high-significance detection of AME in a SNR, which could yet be associated with the nearby HII region G034.7-00.6 that is limited by an annular PDR region containing PAHs (C07). \citet{scaife07} claimed a tentative detection in the SNR 3C396 using data at 33~GHz from the VSA interferometer. However, a reanalysis of these data, in combination with new data from the Parkes-64m telescope between 8 and 19~GHz, indicates that the spectrum of this source is entirely compatible with synchrotron emission, with no need for AME \citep{cruciani16}. In the sample analysed in \citet{pip15} there are some AME regions that could contain some contribution from faint SNRs, but it is not clear what is the real contribution from these SNRs to the total observed flux density, and also these detections are usually at low significance. The confirmation of AME in W44 would require high angular resolution observations in the frequency range from 10 to 30~GHz. They would also be useful to identify the exact location where AME originates, and to elucidate if it is associated with the HII region G034.7-00.6.

 \subsubsection{Dust emission}
The measured FIR flux densities in W44 are indicative of the presence of thermal dust emission which, under the assumption of reliable background removal, may be associated with this object. This hypothesis is consistent with the high-angular resolution data from the {\it Herschel} observatory that have revealed the presence of dust emission in the expanding shell of this SNR\footnote{See {\tt http://sci.esa.int/herschel/51098-annotated\\-composite-image-of-w44/}.}. There has been a longstanding debate about whether or not SN explosions can lead to the formation of dust (see e.g. G\'omez et al. 2012). In any case, here there may also be an important contribution from the nearby HII region G034.7-00.6, which has also been identified in {\it Herschel} data.

\section{Polarised emission from compact sources}\label{sec:polarisation}

Polarisation maps of the Stokes parameters $Q$ and $U$ were shown in Fig.~\ref{fig:large_maps} and in Fig.~\ref{fig:iqu_maps}, and exhibit strong polarisation associated with the compact synchrotron emission of the SNR W44, and with diffuse emission, mainly in $Q$, distributed along the Galactic plane. In section~\ref{sec:diffuse} we found that at low frequency the spectrum of this diffuse polarised signal is compatible with synchrotron emission, whereas the flattening of the spectrum at $\sim 40$~GHz may be explained by polarised thermal dust emission from the Galactic ISM. In this section we study the spectrum of the polarised intensity towards W43, W44 and W47. In Table~\ref{tab:fluxes_qu} we list the flux densities inferred through aperture photometry on the $Q$ and $U$ maps following exactly the same methodology that we applied to the intensity maps in section~\ref{subsec:intensity}. Contrary to intensity, in polarisation the pixel-to-pixel standard deviation in the external annulus is not dominated by background fluctuations but by instrumental noise. For this reason, in polarisation the assumption of noise diagonal covariance matrix does not seem sufficiently conservative due to the presence of correlated $1/f$ noise. Therefore, in this case we resort to a more conservative  estimation consisting of extracting the flux density in 10 apertures located in positions near the sources and calculating the dispersion of these values. These estimates for $Q$ and $U$ are indicated in the second and third columns of Table~\ref{tab:fluxes_qu}, and are found to be typically a factor $2$ to $4$ higher than the previous errors derived from the pixel-to-pixel dispersion in the background annulus. We will use these more conservative estimates in subsequent analyses.

\begin{table*}
\begin{center}
\begin{tabular}{ccccccccccccc}
\hline\hline
\noalign{\smallskip}
  Freq. &&  \multicolumn{2}{c}{Random ap.} && \multicolumn{2}{c}{W43r} && \multicolumn{2}{c}{W44r}  && \multicolumn{2}{c}{W47r} \\
\noalign{\smallskip}
\cline{3-4}\cline{6-7}\cline{9-10}\cline{12-13}
\noalign{\smallskip}
(GHz) && $\sigma_Q$ & $\sigma_U$ && $Q$ (Jy) & $U$ (Jy) &&  $Q$ (Jy) & $U$ (Jy) &&  $Q$ (Jy) & $U$ (Jy)  \\
  \noalign{\smallskip}\hline\noalign{\smallskip}
 1.40  &&  1.35   &  1.65  && $-2.99\pm 0.36$	& $ 5.56\pm 0.39$  && $ 3.20\pm 0.65$	& $20.32\pm 0.70$  && $-5.13\pm 0.24$	& $ 2.12\pm 0.36$   \\
11.1   &&  0.64   &  0.75  &&   --              &   --             && $-6.85\pm 0.21$	& $ 5.58\pm 0.30$  && $ 0.58\pm 0.22$	& $-0.33\pm 0.27$   \\
12.9  &&  0.54   &  0.40  &&   --              &   --             && $-5.12\pm 0.24$	& $ 6.47\pm 0.30$  && $ 0.77\pm 0.18$	& $-2.32\pm 0.19$   \\
16.7  &&  0.46   &  0.44  && $-0.15\pm 0.18$	& $-1.16\pm 0.17$  && $-4.95\pm 0.14$	& $ 5.49\pm 0.19$  && $ 1.44\pm 0.14$	& $-1.07\pm 0.11$   \\
18.7  &&  0.77   &  0.94  && $ 0.20\pm 0.22$	& $ 1.79\pm 0.28$  && $-4.26\pm 0.14$	& $ 5.79\pm 0.18$  && $ 1.00\pm 0.18$	& $ 0.63\pm 0.17$   \\
 22.7  &&  0.20   &  0.23  && $ 0.57\pm 0.12$	& $-0.29\pm 0.09$  && $-2.59\pm 0.13$	& $ 5.72\pm 0.09$  && $ 1.24\pm 0.13$	& $-0.88\pm 0.07$   \\
 28.4  &&  0.17   &  0.19  && $-1.35\pm 0.09$	& $ 1.51\pm 0.08$  && $-2.26\pm 0.09$	& $ 5.47\pm 0.07$  && $ 0.43\pm 0.09$	& $-0.05\pm 0.06$   \\
 32.9  &&  0.20   &  0.19  && $ 0.20\pm 0.09$	& $-0.17\pm 0.06$  && $-1.67\pm 0.10$	& $ 4.94\pm 0.05$  && $ 0.75\pm 0.09$	& $-0.67\pm 0.04$   \\
 40.6  &&  0.17   &  0.16  && $ 0.12\pm 0.07$	& $-0.27\pm 0.05$  && $-1.44\pm 0.09$	& $ 4.46\pm 0.06$  && $ 0.79\pm 0.08$	& $-0.64\pm 0.04$   \\
 44.1  &&  0.15   &  0.21  && $ 0.05\pm 0.06$	& $ 0.56\pm 0.04$  && $-1.07\pm 0.09$	& $ 4.19\pm 0.04$  && $ 0.97\pm 0.06$	& $-0.30\pm 0.06$   \\
 60.5  &&  0.47   &  0.52  && $ 0.22\pm 0.09$	& $ 0.28\pm 0.09$  && $-0.44\pm 0.10$	& $ 3.31\pm 0.06$  && $ 0.15\pm 0.09$	& $-0.37\pm 0.06$   \\
 70.4  &&  0.45   &  0.28  && $ 0.24\pm 0.10$	& $ 0.07\pm 0.05$  && $-0.54\pm 0.14$	& $ 3.12\pm 0.04$  && $ 0.89\pm 0.10$	& $-0.24\pm 0.06$   \\
 93.0  &&  1.39   &  1.40  && $-0.08\pm 0.23$	& $-2.20\pm 0.15$  && $ 1.58\pm 0.31$	& $ 0.38\pm 0.16$  && $ 0.83\pm 0.23$	& $-1.32\pm 0.16$   \\
\noalign{\smallskip}
\hline\hline
\end{tabular}
\normalsize
\caption{Polarisation flux densities for the regions containing W43, W44 and W47, derived from DRAO, QUIJOTE, \wmap and {\it Planck}. They have been calculated through aperture photometry, exactly in the same way as the total-intensity flux densities. The error bars have been calculated through the RMS pixel dispersion in the background annulus. The second and third columns show the corresponding error bars on $Q$ and $U$ derived from the dispersion of the polarisation fluxes calculated in 10 random apertures.}
\label{tab:fluxes_qu}
\end{center}
\end{table*}

In W43r the $Q$ and $U$ flux densities are consistent with zero in most of the cases, whereas in W44r they are dominated by the synchrotron emission of the SNR. Towards W47r we get non-zero polarised flux densities and practically uniform polarisation angles for most of the frequencies, and a spectral index of $-0.56$. This seems to be associated with the diffuse Galactic synchrotron emission that is seen in the maps to the east of W44, extending between $l\approx 35^\circ$ and $l\approx 38^\circ$, rather than to the source itself. Note however that through the correlation plots of Fig.~\ref{fig:pp} we got a somewhat steeper spectral index of $\approx -1.2$ in region 2, which includes this structure between $l\approx 35^\circ$ and $l\approx 38^\circ$. 

In W43r and W47r we detect significant polarisation in the DRAO map at $1.4$~GHz, respectively at $4.0\sigma$ and $4.2\sigma$ and with polarisation angles $\gamma =-59^\circ\pm 8^\circ$ and $\gamma =-79^\circ\pm 11^\circ$. This emission could be explained if these HII regions would be acting as Faraday screens (see e.g. Gao et al. 2010), leading to a rotation of the polarisation angle of background synchrotron emission. This background emission shows up at high frequencies in the case of W47r, with a different polarisation angle than at $1.4$~GHz, although with a low signal to noise that prevent us from obtaining a meaningful estimate of the rotation measure. On the contrary, in the case of W43r the synchrotron emission may have decreased at higher frequencies and be embedded in the noise.

In the following two sections we focus on the constraints on the AME polarisation that can be extracted from our measurements in W43r, and on the characterisation of the polarisation of the synchrotron emission in W44, respectively.

\subsection{Upper limits on the AME polarisation in W43}\label{subsec:ame_pol_w43}

From the $Q$ and $U$ flux densities of Table~\ref{tab:fluxes_qu} we obtain the polarised flux densities, $P=\sqrt{Q^2+U^2}$, listed in Table~\ref{tab:pi_w43}. These quantities have been debiased by integrating the analytical posterior probability density function of the measured polarised flux density \citep{vaillancourt06,rubino12}. Two frequency bands show $P^{\rm db}$ values with signal-to-noise ratios above 2. At $1.4$~GHz the DRAO map shows significant negative emission in $Q$ and positive in $U$ towards W43. As it was explained before, this emission could result from a Faraday-screen effect. The 28.4~GHz \planck map shows negative emission in $Q$, with a spatial structure similar to the free-free emission of the source in total intensity. Conversely, at other frequencies and in particular at 22.7~GHz the emission at this position looks more diffuse and positive in $Q$. Therefore, it seems clear that the 28.4~GHz point is affected by residual intensity to polarisation leakage at this position. As it was discussed in section~\ref{sec:ancillary_data} we applied the bandpass-mismatch correction maps. With respect to total intensity, they introduce corrections of the order of $2-4\%$ at $28.4$~GHz, $\sim 1\%$ at $70.4$~GHz, and much smaller at $44.1$~GHz. However they are only reliable in scales larger than $\sim 1^\circ$, and this may explain the residual leakage at the position of the sources. For this reason, we will not use any {\it Planck} polarisation data in all subsequent analyses.

On the other hand, the emission detected at 22.7~GHz seems to be real, and comes mainly from the positive signal that shows up in the $Q$ map. This signal, even though at a level similar to the diffuse emission that is seen along the Galactic plane, becomes slightly more intense inside the aperture, at a position matching the central coordinates of W43. This leads us to consider the possibility of this polarisation being intrinsic to the source, in which case it would inevitably be associated either with the free-free or with the AME. The free-free emission from a Maxwellian distribution of electrons is known to be practically unpolarised, with the possibility of some residual net polarisation, below $\sim 1\%$, originated in the borders of the region \citep{trujillo02}. The polarisation fraction of our signal with respect to the free-free intensity at 22.7~GHz is $P/I_{\rm ff}\times 100\approx 0.3\%$, and therefore compatible with the previous level. The characterisation of the spectrum would be helpful to elucidate whether this polarisation is associated with the AME or with the free-free emission of the source or, alternatively, with the diffuse Galactic synchrotron emission. Unfortunately, this is not possible given the level of the noise at the other frequencies. Our measurements between 17 and 33~GHz are compatible with any of these mechanisms. Note that if the responsible mechanism for this emission were synchrotron, it would have a very high polarisation or a very hard spectrum in order to make the observed polarisation compatible with the fact that the intensity SED in W43r does not show evidence of any synchrotron component. In fact, assuming a typical polarisation degree of $10\%$, the total-intensity flux density at 22.7~GHz would be 7.7~Jy. Extrapolating this value to 0.408~GHz with a typical spectral index $\beta_{\rm sync}=-1$ would give 428~Jy, which should be detectable in the intensity SED. We have tried to fit the SEDs of W43r and W47r including a synchrotron component, but this resulted in a poorer $\chi^2$.

\begin{table}
\begin{center}
\begin{tabular}{cccccc}
\hline\hline
\noalign{\smallskip}
  Freq. &&  \multicolumn{3}{c}{W43r} \\
\noalign{\smallskip}
\cline{3-5}
\noalign{\smallskip}
(GHz)  &&  $I_{\rm AME}$ & $P^{\rm db}$ & $\left(P/I_{\rm AME}\right)^{\rm db}$  \\
       && (Jy)           & (Jy)         &  $\times 100$ (\%)    \\
  \noalign{\smallskip}\hline\noalign{\smallskip}
 1.40  &&  $-$          &  $6.31\pm1.59$          &  $-$                    \\
\noalign{\smallskip}
16.7  &&  $241\pm 12$  &  $<0.93$                &  $<0.39$                \\
\noalign{\smallskip}
18.7  &&  $269\pm 13$  &  $<1.93$                &  $<0.71$                \\
\noalign{\smallskip}
22.7  &&  $238\pm 16$  &  $0.77\pm 0.23$         &  $0.32\pm 0.10$         \\
\noalign{\smallskip}
32.9   &&  $224\pm 15$  &  $0.10^{+0.21}_{-0.10}$ &  $<0.24$                \\
\noalign{\smallskip}
40.6   &&  $186\pm 14$  &  $<0.40$                &  $<0.22$                \\
\noalign{\smallskip}
44.1   &&  $172\pm 14$  &  $<0.43$                &  $<0.25$                \\
\noalign{\smallskip}
60.5   &&  $118\pm 16$  &  $<1.14$                &  $<0.98$                \\
\noalign{\smallskip}
70.4   &&  $107\pm 21$  &  $<0.74$                &  $<0.73$                \\
\noalign{\smallskip}
93.0   &&  $ 92\pm 48$  &  $<2.81$                &  $<3.12$                \\
\noalign{\smallskip}
\hline\hline
\end{tabular}
\normalsize
\caption{AME residual flux densities, polarised flux densities and polarisation fractions for W43r. The residual flux densities have been calculated by subtracting the free-free and thermal dust modelled flux densities to the measurements shown in Table~\ref{tab:fluxes}. The polarisation fractions are referred to these values. The error bars represent $1\sigma$ uncertainties. In their calculation we have assumed for the $Q$ and $U$ errors the values shown in the second and third columns of Table~\ref{tab:fluxes_qu}. Upper limits are referred to the 95\% C.L.}
\label{tab:pi_w43}
\end{center}
\end{table}

In Table~\ref{tab:pi_w43} we also show the polarisation fractions relative to the AME residual flux densities in intensity, $I_{\rm AME}=I^{\rm meas}-I_{\rm ff}^{\rm mod}-I_{\rm dust}^{\rm mod}$, where $I^{\rm meas}$ is the measured flux density in total intensity shown in Table~\ref{tab:fluxes}, and $I_{\rm ff}^{\rm mod}$ and $I_{\rm dust}^{\rm mod}$ are the free-free and thermal dust flux densities derived from our fitted models (see section~\ref{subsec:intensity_w43_w44_w47} and Fig.~\ref{fig:seds}). As we did in section~\ref{sec:intensity} for total intensity, in this analysis we only include statistical uncertainties in the error bar of the polarised flux densities. In general, and in particular in QUIJOTE, where we apply the same calibration strategy in polarisation and in intensity, any potential calibration error will cancel out in the ratio $P^{\rm meas}/I^{\rm meas}$. Therefore, the polarisation fractions $\Pi_{\rm AME}$ would only be affected by calibration errors through the modelled intensity flux densities.

In order to debias the quantity $\Pi_{\rm AME}$ we integrate a probability density function that has been calculated numerically through Monte Carlo simulations, in the same way it was done in \citet{genova15}. The upper limits here are referred to the 95\% C.L. The measured polarisation fraction at 22.7~GHz, $\Pi=0.32\pm 0.10\%$, is still compatible with the previous upper limits in the literature: $\Pi_{\rm AME}<1\%$, obtained at the same frequency in G159.6-18.5 within the Perseus molecular complex \citep{lopez11,dickinson11}. It was argued in the previous paragraph that this polarised signal could be associated with the background synchrotron or even with the free-free emission of the source. Even if we cannot interpret this as a detection of AME polarisation, using this measurement we can still set an upper limit of $\Pi_{\rm AME}<0.52\%$ at the 95\% C.L., better by a factor 2 than previous constraints. Using the \wmap measurement at 40.6~GHz we get an even stronger upper limit of $\Pi_{\rm AME}<0.22\%$, which is better than previous constraints by almost a factor 5. The most stringent QUIJOTE upper limit is $\Pi_{\rm AME}<0.39\%$, at 16.7~GHz.

\begin{figure*}
\centering
\includegraphics[width=12cm]{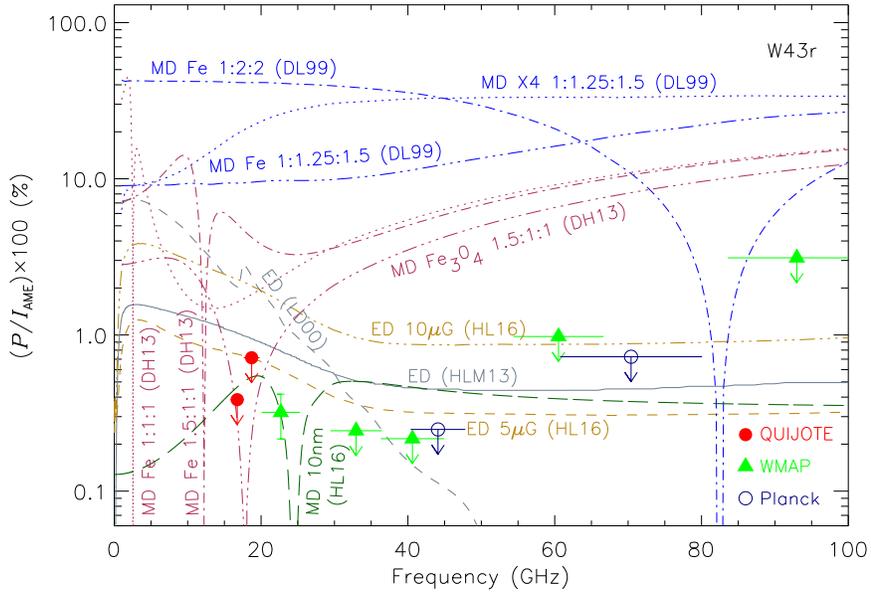}
\caption{\small Constraints on several microwave emission models for the electric dipole (ED) emission and the magnetic dipole (MD) emission based on our measurements on W43r. We plot the ratio between the polarised flux density measured at the position of the W43 HII region and the AME residual flux density (measured flux density in total intensity minus the combination of free-free and thermal dust emissions). The \wmap value at 22.7~GHz may be contaminated by background diffuse emission. At all other frequencies the observed polarisation is consistent with zero and we represent upper limits at the 95\% C.L. The horizontal lines around each data point represent the bandwidth of the corresponding detector. Our measurements are compared with theoretical models predicting the frequency dependence of the fractional polarisations of the ED and MD emissions (see text for details).}
\label{fig:pol_constraints_w43}
\end{figure*}

These constraints on the fractional polarisation with respect to the AME residual flux density, $\Pi_{\rm AME}$, are plotted in Fig.~\ref{fig:pol_constraints_w43}, except the 28.4~GHz value which have been ignored due to being affected by leakage. This plot is an update of Fig.~8 of \citet{genova15}, where we showed the upper limits derived on G159.6-18.5, together with previous constraints in the literature in this and other regions. Our constraints on W43r are compared in Fig.~\ref{fig:pol_constraints_w43} with theoretical predictions for the polarisation fraction of the electric dipole (ED) emission (Lazarian \& Draine 2000, hereafter LD00; Hoang et al. 2013, hereafter HLM13; Hoang \& Lazarian 2015, hereafter HL16)\footnote{Note that in Fig.~8 of \citet{genova15} the HLM13 model for the electric dipole emission had polarisation fractions a factor $\approx 2$ higher than in Fig.~\ref{fig:pol_constraints_w43} of this paper. The reason for this is that in \citet{genova15} we relied on the model of a previous arXiv version of the HLM13 paper (arXiv:1305.0276v1), whereas here we show the model in the final published version, which is different.}. In the case of HL16 we plot their estimates for grain temperature $T_{\rm d}=60$~K, and for two values of the magnetic field intensity, $5$ and $10~\mu$G. They also show polarisation fractions for $T_{\rm d}=40$~K, which are a factor $\sim 2$ higher, and for stronger (probably unrealistic) magnetic fields, which also result in higher polarisation fractions. 

All the ED models represented in Fig.~\ref{fig:pol_constraints_w43} are clearly inconsistent with our data. Note in particular how our upper limit from QUIJOTE at 16.7~GHz, $<0.39\%$ (95\% C.L.), compares with different ED models at the same frequency: $1.94\%$ from LD00, $1.03\%$ from HLM13, and $0.75\%$ and $2.3\%$ from HL16 for $5~\mu$G and $10~\mu$G, respectively. It is convenient to note however that the LD00 and the HLM13 models represent upper limits on the real spinning dust polarisation. LD00 considered resonance relaxation as the mechanism responsible for grain alignment, but disregarded saturation effects that may lead to lower alignment efficiencies and in turn lower polarisation fractions. On the other hand, HLM13 inferred the efficiency of the grain alignment through observations of the UV polarisation bump, assuming that only polycyclic aromatic hydrocarbon (PAH) molecules, those which are thought to generate the spinning dust emission, are responsible for this polarisation bump. However, if graphite grains would also contribute, then the efficiency of the alignment of PAHs would actually be lower and so would be the degree of spinning dust polarisation. Given that our upper limits are typically at least a factor $2$ below the model, we understand that a considerable contribution from graphite grains would be needed in order to accommodate the model with our data. Similarly, lower magnetic field strengths or higher grain temperatures could make the HL16 model compatible with our measurements, even though at the cost of less realistic physical conditions. In a more recent paper, \citet{draine16} proposed that the quantisation of energy levels in very small grains may result in a suppression of energy dissipation which in turn would lead to a dramatic decrease of the alignment efficiency. At frequencies $>10$~GHz small grains ($<10\AA$) would show practically no polarisation ($<10^{-4}\%$), and only bigger grains ($>30\AA$) would produce polarised emission at a level above our derived upper limits.

We also plot in Fig.~\ref{fig:pol_constraints_w43} models predicting the polarisation fractions associated with the magnetic dipole (MD) emission (Draine \& Lazarian 1999, hereafter DL99; Draine \& Hensley 2013, hereafter DH13; HL16). DL99 considered dust grains ordered in a single magnetic domain, and in this case we plot the models for different grains geometries (spheroids, with different axial ratios $a_1$:$a_2$:$a_3$, as indicated in the figure) and compositions (also indicated next to each curve). Except near the crossover frequency, where the polarisation sign flips (note that we plot absolute values of the polarisation fraction), all these models predict polarisation levels of up to $10\%$ and are ruled out by our data. DH13 considered the magnetic particles to be inclusions randomly oriented inside the grains, which causes the polarisation to decrease. We show in Fig.~\ref{fig:pol_constraints_w43} three different combinations of grain shapes and compositions, including magnetite (Fe$_3$O$_4$) with spheroidal shape ($1.5$:$1$:$1$) which, of all cases presented in DH13, is the one predicting the lowest polarisations. Although not with QUIJOTE, which lies close to the crossover frequency, this model is inconsistent with the upper limits coming from \wmap and \planck data. It must be noted however that DH13 considered perfect alignment between the grain angular momentum and the magnetic field. If they would not be perfectly aligned the polarisation fraction would be notably reduced (see Fig.~9 of DH13). Recently HL16 presented a more exhaustive study of MD emission from free-flying magnetic nanoparticles, considering different sizes, grain temperatures, and two magnetic susceptibilities (DL99 and DH13), and in some cases find lower polarisation fractions. In Fig.~\ref{fig:pol_constraints_w43} we represent the HL16 model with the lowest polarisation levels, which corresponds to grains of 10~nm, $T_{\rm d}=40$~K and the magnetic susceptibility of DH13. Even this model is inconsistent with the QUIJOTE 16.7~GHz point, with \wmap 22.7~GHz, 32.9~GHz and 40.6~GHz, and with \planck 44.1~GHz. Smaller grains (they consider sizes down to 0.55~nm) are more efficiently aligned and give higher polarisations. On the contrary, hotter grains are expected to result in lower polarisations. However, the maximum temperature they consider is $T_{\rm d}=40$~K, which is what we plot.

It is therefore clear from Fig.~\ref{fig:pol_constraints_w43} that, unless we invoke quantum dissipation of alignment \citep{draine16}, or misalignment between the grains and the magnetic field or between the magnetic field and the plane of the sky, there is not any single model compatible with our constraints, which are the most stringent obtained to date on the AME polarisation. This may have important implications on the theoretical models of spinning dust emission, so we have to consider various aspects which could make our data compatible with the models. One possibility is that the AME would have polarisation direction orthogonal to the diffuse synchrotron in the direction of W43, or alternatively to the hypothetical net polarisation associated with the free-free emission of the source itself. However in order for the two components to cancel their polarisations, not only orthogonality would be needed but also equal amplitudes. This cannot happen at all frequencies given the difference of the AME spectrum with respect to the two other components, so we should expect some non-zero polarisation at some frequencies. This occurs at 22.7~GHz but, yet, it seems difficult to imagine a combination of polarisation angles and amplitudes of the two components that could produce the observed behaviour. 

In section~\ref{subsec:intensity} we argued that, owing to the coarse angular resolution and to the location of the sources, our flux densities could be subject to significant background contamination. In any case, the actual origin of the AME do not have implications on the constraints on the AME polarisation that we discuss here. It is more critical to have an accurate modelling of the SED, as this will result in a correct estimation of the residual AME. In the case of W43r, this is strongly reliant on having a correct estimation of the level of free-free emission, that is essentially pinned down by the low-frequency data. Fig.~\ref{fig:seds} demonstrates that the fitted model is in good agreement with these data. However, as it was discussed in section~\ref{subsec:intensity_w43_w44_w47}, our best-fit free-free amplitude is marginally below the expectation from RRL data. We get $EM=3911\pm 68$~cm$^{-6}$~pc, whereas the RRL survey of \citet{alves12} predicts a value of the emission measure in the range $4020-6190$~cm$^{-6}$~pc. If we consider as a reference the central value of this interval, the amplitude of the free-free emission would increase by 30\%. At 40.6~GHz, the frequency of our most stringent constraint, this would result in a decrease of the residual AME that would leave the upper limit on the AME polarisation at $\Pi_{\rm AME}<0.40\%$, a value that is yet considerably smaller than previous constraints. Nonetheless, in this case the four data points between $0.408$~GHz and $2.33$~GHz would be respectively $-30\sigma$, $-33\sigma$, $-37\sigma$ and $-28\sigma$ away from the model. Given the level of consistency of these low-frequency data points, it seems more reasonable to rely on them rather than on the RRL observations to fix the level of free-free. Finally, it is also important to have a reliable characterisation of the spinning dust in intensity. As it was discussed in section~\ref{subsec:intensity_w43_w44_w47}, the best-fit spectrum seems too broad in comparison with physically-motivated spinning dust models. This needs to be understood.

\begin{table}
\begin{center}
\begin{tabular}{cccc}
\hline\hline
\noalign{\smallskip}
  Freq. &&  \multicolumn{2}{c}{W44} \\
\noalign{\smallskip}
\cline{3-4}
\noalign{\smallskip}
(GHz) &&  $Q$ (Jy) & $U$ (Jy)  \\
  \noalign{\smallskip}\hline\noalign{\smallskip}
 1.40  && $ 3.70\pm 0.41$   & $16.28\pm 0.44$  \\  
11.1   && $-9.21\pm 0.21$   & $ 6.23\pm 0.30$  \\  
12.9  && $-8.04\pm 0.24$   & $ 7.31\pm 0.30$  \\  
16.7  && $-6.39\pm 0.14$   & $ 5.31\pm 0.19$  \\  
18.7  && $-5.93\pm 0.14$   & $ 6.45\pm 0.18$  \\  
 22.7  && $-3.92\pm 0.08$   & $ 6.06\pm 0.05$  \\  
 28.4  && $-3.21\pm 0.06$   & $ 5.47\pm 0.04$  \\  
 32.9  && $-2.61\pm 0.06$   & $ 4.92\pm 0.03$  \\  
 40.6  && $-2.44\pm 0.06$   & $ 4.54\pm 0.03$  \\  
 44.1  && $-2.04\pm 0.05$   & $ 4.44\pm 0.02$  \\  
 60.5  && $-1.43\pm 0.06$   & $ 3.63\pm 0.04$  \\  
 70.4  && $-2.01\pm 0.09$   & $ 3.27\pm 0.03$  \\  
 93.0  && $-1.49\pm 0.19$   & $ 1.54\pm 0.10$  \\  
\noalign{\smallskip}
\hline\hline
\end{tabular}
\normalsize
\caption{Polarisation flux densities for W44, derived from DRAO, QUIJOTE, \wmap and {\it Planck}. They have been calcualted by integrating all pixels in a $1^\circ$ radius and subtracting a background level given by the emission profile defined by two constant-longitude cuts at $l_{\rm c1}=33.2^\circ$ and $l_{\rm c2}=36.2^\circ$.}
\label{tab:fluxes_qu_w44}
\end{center}
\end{table}

\subsection{Polarised synchrotron emission from W44}\label{subsec:sync_pol_w44}

The $Q$ and $U$ fluxes of W44r shown in Table~\ref{tab:fluxes_qu} were obtained through aperture photometry, with an aperture of radius 60~arcmin and subtracting a median background level calculated in an external ring between 80 and 100~arcmin, exactly the same procedure that was applied in section~\ref{subsec:intensity} on the intensity maps. With the goal of better isolating the source emission from the diffuse background, in section~\ref{subsec:intensity_w44} we applied a different strategy consisting in defining the background level through two cuts at constant Galactic longitudes. In Table~\ref{tab:fluxes_qu_w44} we show the polarisation flux densities of W44 obtained through the same procedure. Differences with respect to the values of Table~\ref{tab:fluxes_qu} are typically below $1\sigma$ for $U$, and of the order of $2-4\sigma$ for $Q$, still lower than the differences found in intensity thanks to the weaker impact of the background emission in the polarisation data.
 
\begin{table*}
\begin{center}
\begin{tabular}{cccccccccccccc}
\hline\hline
\noalign{\smallskip}´
  & \multicolumn{6}{c}{W44r} && \multicolumn{6}{c}{W44} \\
\noalign{\smallskip}
\cline{2-7}\cline{9-14}
\noalign{\smallskip}
  && \multicolumn{2}{c}{With free-free} && \multicolumn{2}{c}{Without free-free} &&& \multicolumn{2}{c}{With free-free} && \multicolumn{2}{c}{Without free-free} \\
\noalign{\smallskip}
\cline{3-4}\cline{6-7}\cline{10-11}\cline{13-14}
\noalign{\smallskip}
Freq.  &  $P^{\rm db}$ & $I_{\rm sync}$ & $\Pi_{\rm sync}^{\rm db}$ && $I_{\rm sync}$ & $\Pi_{\rm sync}^{\rm db}$ &&$P^{\rm db}$ & $I_{\rm sync}$ & $\Pi_{\rm sync}^{\rm db}$ && $I_{\rm sync}$ & $\Pi_{\rm sync}^{\rm db}$ \\ 
(GHz) &  (Jy)  &  (Jy) & (\%) && (Jy)  &  (\%) &&(Jy)  &  (Jy) & (\%) && (Jy)  &  (\%)  \\
 \noalign{\smallskip}\hline\noalign{\smallskip}
     1.40 &	 $21\pm 2$     &$193\pm 12$ &  $11\pm 1$ &&$329\pm 12$   &  $6.3\pm 0.5$ &&	  $17\pm 2$    & $107\pm 7$  &        $16\pm 2$ && $197\pm 7$  &    $8.5\pm 0.9$ \\
  11.1  &	$8.8\pm 0.7$  &  $46\pm 7$ &  $19\pm 3$ && $136\pm 8$   &  $6.5\pm 0.6$ &&	  $11\pm 1$    &  $22\pm 7$  &       $50\pm 16$ &&  $96\pm 7$  &     $12\pm 1$ \\
  12.9 &	$8.3\pm 0.5$  &  $53\pm 8$ &  $16\pm 2$ && $141\pm 9$   &  $5.8\pm 0.5$ &&	  $11\pm 0.5$  &  $23\pm 7$  &       $47\pm 14$ &&  $96\pm 7$  &     $11\pm 1$ \\
  16.7 &	$7.4\pm 0.4$  &  $39\pm 8$ &  $19\pm 4$ &&$125\pm 11$   &  $5.9\pm 0.6$ &&	 $8.3\pm 0.4$  &  $17\pm 7$  &       $50\pm 21$ &&  $85\pm 7$  &    $9.8\pm 1.0$ \\
  18.7 &	$7.2\pm 0.9$  &  $45\pm 9$ &  $16\pm 4$ &&$129\pm 11$   &  $5.6\pm 0.8$ &&	 $8.8\pm 0.9$  &  $17\pm 7$  &       $52\pm 23$ &&  $82\pm 8$  &     $11\pm 1$ \\
  22.7  &	$6.3\pm 0.2$  & $25\pm 11$ & $26\pm 12$ &&$104\pm 14$   &  $6.0\pm 0.8$ &&	 $7.2\pm 0.2$  &  $13\pm 7$  &       $57\pm 30$ &&  $73\pm 7$  &    $9.9\pm 1.0$ \\
  28.4  &	$5.9\pm 0.2$  & $27\pm 11$ &  $22\pm 9$ &&$100\pm 13$   &  $5.9\pm 0.8$ &&	 $6.3\pm 0.2$  &   $9\pm 6$  &       $69\pm 48$ &&  $65\pm 7$  &    $9.8\pm 1.1$ \\
  32.9  &	$5.2\pm 0.2$  & $26\pm 10$ &  $20\pm 8$ && $94\pm 12$   &  $5.6\pm 0.7$ &&	 $5.6\pm 0.2$  &  $13\pm 6$  &       $42\pm 18$ &&  $67\pm 6$  &    $8.3\pm 0.8$ \\
  40.6  &	$4.7\pm 0.2$  &  $24\pm 8$ &  $19\pm 7$ && $87\pm 10$   &  $5.4\pm 0.6$ &&	 $5.2\pm 0.2$  &  $10\pm 5$  &       $49\pm 24$ &&  $63\pm 5$  &    $8.2\pm 0.7$ \\
  44.1  &	$4.3\pm 0.2$  &  $25\pm 8$ &  $18\pm 6$ &&  $87\pm 9$   &  $5.0\pm 0.6$ &&	 $4.9\pm 0.2$  &  $10\pm 5$  &       $47\pm 22$ &&  $62\pm 5$  &    $7.8\pm 0.7$ \\
  60.5  &	$3.3\pm 0.5$  &  $21\pm 8$ &  $16\pm 7$ &&  $82\pm 9$   &  $4.1\pm 0.8$ &&	 $3.9\pm 0.5$  &   $6\pm 5$  &       $69\pm 60$ &&  $56\pm 6$  &    $6.9\pm 1.1$ \\
  70.4  &	$3.2\pm 0.3$  & $22\pm 10$ &  $14\pm 7$ && $83\pm 12$   &  $3.8\pm 0.7$ &&	 $3.8\pm 0.3$  &   $4\pm 6$  &      $90\pm 123$ &&  $54\pm 7$  &    $7.2\pm 1.2$ \\
  93.0  &$1.4^{+1.1}_{-1.4}$  & $22\pm 20$ &	  $<32$ && $79\pm 26$ &	 $<6.2$ &&$0.9^{+1.4}_{-0.9}$  &  $5\pm 11$  &         $<226$ && $49\pm 16$  &    	   $<9.4$ \\ 
\noalign{\smallskip}
\hline\hline
\end{tabular}
\normalsize
\caption{Measured polarised flux densities for W44r and W44, and derived polarisation fractions with respect to the residual modelled synchrotron intensities, $\Pi_{\rm sync}^{\rm db}=\left(P/I_{\rm sync}\right)^{\rm db}\times 100$. We show the results when we use the estimates of $I_{\rm sync}$ coming from the modelling of the intensity SED with and without including a free-free component. For W44, we have considered the model without free-free that represents the best fit to the combination of our and C07 data points. The error bars represent $1\sigma$ uncertainties, whereas upper limits are referred to the 95\% C.L.}
\label{tab:pi_w44}
\end{center}
\end{table*}

\begin{figure*}
\centering
\includegraphics[width=15cm]{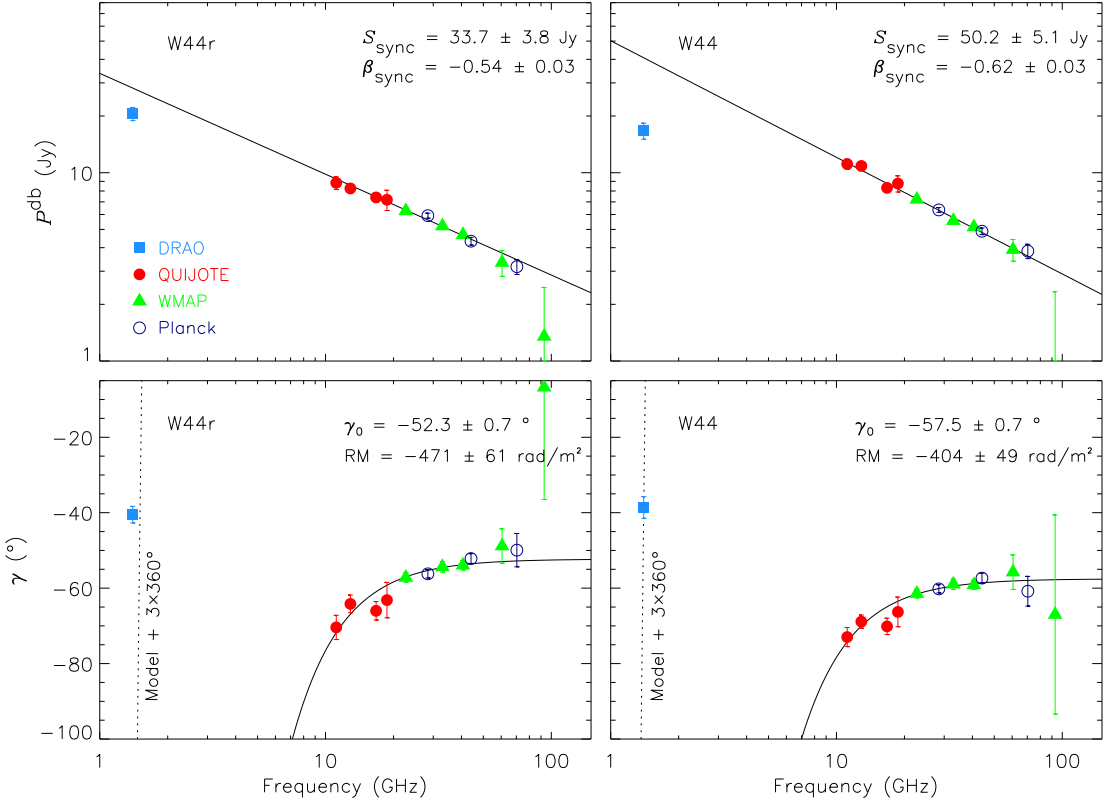}
\caption{\small Spectrum of the polarised flux density (top) and dependence of the polarisation direction with the frequency (bottom) for W44r (left) and W44 (right). The polarisation spectrum is fitted to a synchrotron power-law, whereas the dependence of the polarisation angle with the frequency is fitted with a Faraday rotation model. The DRAO point is not included in the fit because it looks affected by Faraday depolarisation. The best-fit parameters are shown in the top-right corner of each panel.}
\label{fig:sed_pol_w44}
\end{figure*}

\begin{table}
\begin{center}
\begin{tabular}{ccc}
\hline\hline
\noalign{\smallskip}
Parameter &  W44r & W44 \\
\noalign{\smallskip}\hline\noalign{\smallskip}                
      $S_{\rm sync}^{\rm 1 GHz}$ (Jy)  &       $  33.7\pm  3.8$       &    $  50.2\pm  5.1$        \\
\noalign{\smallskip}   
                 $\beta_{\rm sync}$  &       $ -0.54\pm 0.03$       &    $ -0.62\pm 0.03$        \\
\noalign{\smallskip}   
                 $\chi^2_{\rm red}$  &       $      0.82    $       &    $      0.91    $        \\
\noalign{\smallskip}\hline\noalign{\smallskip}
              $\gamma_0$ ($^\circ$)  &       $ -52.3\pm  0.7$       &    $ -57.5\pm  0.7$        \\
\noalign{\smallskip}   
                     RM (rad/m$^2$)  &       $  -471\pm   61$       &    $  -404\pm   49$        \\
\noalign{\smallskip}   
                 $\chi^2_{\rm red}$  &       $      1.33    $       &    $      1.00    $        \\
\noalign{\smallskip}
\hline\hline
\end{tabular}
\normalsize
\caption{Model parameters and reduced chi-squared resulting from the fit of the polarisation spectra of W44r and W44. We separately fit: i) the spectrum of the polarised flux density to a synchrotron power-law (above the line) and ii) the frequency dependence of the polarisation angle to a model with Faraday rotation (below the line). These two models are represented in the top and bottom panels of Fig.~\ref{fig:sed_pol_w44}, respectively.}
\label{tab:parameters_w44_pol}
\end{center}
\end{table}

\subsubsection{Polarisation SED}

Table~\ref{tab:pi_w44} shows the corresponding debiased polarised flux densities for W44r and W44 which are dominated by the synchrotron emission of the SNR. These values are fitted to a pure synchrotron power-law, as illustrated in Fig.~\ref{fig:sed_pol_w44}. When we ignore in this fit the DRAO $1.4$~GHz data point, which looks affected by Faraday depolarisation, the resulting model traces pretty well the observed QUIJOTE, \wmap and \planck data, with reduced chi-squared of $0.82$ and $0.91$ respectively for W44r and W44 (see Table~\ref{tab:parameters_w44_pol}). One caveat is that the \wmap $93.0$~GHz point seems too low, being $-1.4\sigma$ away from the model for W44r, and $-1.5\sigma$ for W44. The polarised emission in this map is consistent with zero, but given the noise it seems it should be detected, at least marginally. 

Interestingly, the best-fit spectral indices are similar to the ones resulting from the intensity SED when the free-free component is introduced in the fit. In this case, the differences between the intensity and polarisation indices are just $-1.4\sigma$ and $-2.0\sigma$, respectively for W44r and W44. On the other hand, when no free-free emission is considered, the fitted synchrotron spectrum in intensity becomes much flatter, and in this case the discrepancies are at $+5.1\sigma$ and $+5.8\sigma$ ($+8.5\sigma$ when the C07 points are introduced in the fit), respectively. This goes in favour of the presence of a free-free component in the total-intensity emission of W44. In sections~\ref{subsec:intensity_w43_w44_w47} and \ref{subsec:intensity_w44} we discussed other favourable arguments, in particular the $\gamma$-ray spectrum discussed in \citet{cardillo14}.
 
\subsubsection{Polarisation fraction}

Note that, contrary to what happens in intensity, in polarisation the flux densities are higher for W44 than for W44r. The main reason for this is the characteristic distribution of the background emission in the $Q$ map around W44, with a negative signal at the position of the source, and strongly positive towards the east. This positive signal is more effectively captured by the Galactic cut at $l_{\rm c2}=36.2^\circ$, shifting the $Q$ flux density to more negative values in W44 and, as a result, boosting the inferred polarisation fraction. 

If we compare the best-fit synchrotron amplitudes at 1~GHz in polarisation and in intensity (with a free-free component) we get $\Pi_{\rm sync}^{\rm 1 GHz}=15.2\pm 1.8\%$ for W44r and $\Pi_{\rm sync}^{\rm 1 GHz}=35.9\pm 3.8\%$ for W44. This value is a bit high, but still possible in the case of highly ordered magnetic fields. \citet{battye11} performed a statistical analysis of the polarisation properties of sources in the \wmap catalogue and found a mean polarisation fraction of $\approx 3.5\%$, with few sources reaching $25\%$. It must be noted however that these are polarisation fractions with respect to the total emission in intensity, so once the other components are subtracted the real polarisation fraction of the synchrotron could be higher. Using observations at 10.7~GHz, \citet{kundu72} found polarisation fractions of $\sim 20\%$ towards the northeast of the SNR, but lower values in the rest of the region. We would then expect the integrated polarisation fraction to be below $20\%$. This is however the polarisation fraction with respect to the total emission at that frequency. At the QUIJOTE frequency of $11.1$~GHz we measure in W44 a total flux density of 117~Jy, so the polarisation fraction with respect to the total emission would be $9.4\%$ at this frequency, a value that could be consistent with \citet{kundu72}. If we use as reference for W44 the fitted synchrotron amplitudes at 1~GHz when no free-free component is introduced in the fit we get $\Pi_{\rm sync}^{\rm 1 GHz}=21.1\pm 2.2\%$, or $\Pi_{\rm sync}^{\rm 1 GHz}=21.9\pm 2.2\%$. These polarisation fractions seem more typical of synchrotron emission.

We also show in Table~\ref{tab:pi_w44}, at each frequency, the resulting polarisation fractions, $\Pi_{\rm sync}^{\rm db}=\left(P/I_{\rm sync}\right)^{\rm db}\times 100$, relative to the synchrotron residual flux densities in intensity, which are calculated by subtracting from the observed values of Table~\ref{tab:fluxes} the different model components fitted in sections~\ref{subsec:intensity_w43_w44_w47} and \ref{subsec:intensity_w44}: $I_{\rm sync}=I^{\rm meas}-I_{\rm ff}^{\rm mod}-I_{\rm AME}^{\rm mod}-I_{\rm dust}^{\rm mod}$. Note that in the case with free-free the polarisation fractions calculated at each frequency are always consistent with the previous values obtained from the ratio of the fitted $S_{\rm sync}^{\rm 1GHz}$. When we consider the model with no free-free emission, the resulting polarisation fractions are of course lower.

\subsubsection{Polarisation angle}

In the bottom panels of Fig.~\ref{fig:sed_pol_w44} we represent the polarisation angles measured in W44r and in W44 versus the frequency. The rotation of the polarisation direction seen at low frequencies is characteristic of Faraday rotation produced by the intervening Galactic magnetic field between the source and the observer. We fit this effect to a model $\gamma=\gamma_0+{\rm RM}\times \lambda^2$, where $\gamma_0$ is the polarisation direction at $\lambda=0$ ($\nu\rightarrow\infty$ in our case) and RM represents the rotation measure. We remove the DRAO data point from this fit in order to avoid the uncertainty associated with the number of turns of the polarisation direction between 1.4~GHz and 11~GHz. The solid lines in Fig.~\ref{fig:sed_pol_w44} represent the best-fit models. The dotted lines, which are obtained adding a constant factor of $3\times 360^\circ$ to these models, are in reasonable agreement with the DRAO point, especially in W44, and show that the polarisation angle has rotated a little less than 3 turns between 1.4 and $11.1$~GHz.

The best-fit model parameters are shown in Table~\ref{tab:parameters_w44_pol}, together with the reduced chi-squared that are very close to unity. Our fitted values of the RM seem a bit high. \citet{oppermann12} have produced a low-resolution ($\approx 0.5^\circ$) full-sky map of the Faraday rotation\footnote{We have downloaded this map, in \healpix projection and with pixel resolution $N_{\rm side}=128$, from {\tt http://wwwmpa.mpa-garching.mpg.de/ift/faraday/2012/index.html}.}. Along the Galactic plane they find on average $|{\rm RM}|\sim 100$~rad~m$^{-2}$, although in some specific regions their map exhibits extreme values below $-1200$~rad~m$^{-2}$ or above $1400$~rad~m$^{-2}$. At the position of W44 they find a mean value of $\sim -250$~rad~m$^{-2}$. \citet{kundu72} produced a RM map of this SNR, finding strong spatial variations in a range between $-330$~rad~m$^{-2}$ and $+70$~rad~m$^{-2}$, and with a mean value of $-92$~rad~m$^{-2}$, considerable lower than our integrated value. More recently \citet{sun11} found values of $-55$~rad~m$^{-2}$ and $-105$~rad~m$^{-2}$ towards the southern and northern parts of W44, respectively, and $-140$~rad~m$^{-2}$ at the position of the pulsar PSR~J1856+0113, located at $(l,b)$ = $(34.56^\circ,-0.5^\circ)$.

\section{Conclusions}\label{sec:conclusions}

We have presented new QUIJOTE data in intensity and in polarisation, in four individual frequencies between 10 and 20~GHz, covering a $\approx 300$~deg$^2$ region of the Galactic plane between $\l\sim 24^\circ$ and $l\sim 45^\circ$, and resulting from 210~h of dedicated observations. The QUIJOTE maps show, for the first time, compact and diffuse polarisation, demonstrating the potential of this experiment to obtain precise measurements of the anisotropies in the CMB polarisation, as well as of the diffuse low-frequency foregrounds, which constitute the two objectives for which it was conceived. With an angular resolution of $\approx$1$^\circ$, these maps have sensitivities in $Q$ and $U$ between 30 and $80~\mu$K~beam$^{-1}$, depending on the frequency, which are roughly consistent with the instrument nominal sensitivities given the integration times.

The QUIJOTE maps reveal diffuse polarised intensity distributed along the Galactic plane, with an orientation perpendicular to the direction of the Galactic magnetic field (positive $Q$ flux, and $U$ close to zero). In combination with other intensity and polarisation surveys, including \wmap and {\it Planck}, we perform correlation analyses to study the spectral properties in intensity and in polarisation of this emission, in two rectangular regions defined by $|b|<2^\circ$ and $25.7^\circ<l<33.7^\circ$ and $35.8^\circ<l<43.8^\circ$. In intensity we find temperature spectral indices of $\sim -0.1$, characteristic of free-free emission, with a flattening between 11 and 19~GHz and a steepening between 19 and 23~GHz, a behaviour that is consistent with having diffuse AME with a peak frequency close to 19~GHz. In polarisation we obtain spectral indices of $\sim -1.2$, which are consistent with synchrotron emission, with a flattening at frequencies above 33~GHz, something that could be indicative of a contribution from polarised thermal dust emission.

The observed area enclose, among other sources, the two molecular complexes W43 and W47, and the SNR W44. Using QUIJOTE data, which are crucial to trace the downturn of the spectrum at frequencies $\lesssim$20~GHz predicted by models based on electric dipole emission from spinning dust grains, we confirm the presence of AME in these three regions. We argue however that care must be taken because the coarse angular resolution of the data may result in important contamination from the diffuse foreground emission. We then use the notation W43r, W44r and W47r, to emphasise that we could be studying the properties of a wider region than the sources themselves. Using a more careful background subtraction in W44, we manage to better isolate the emission from this source, and obtain low-frequency flux densities that agree with the high-angular resolution data of C07.

At the QUIJOTE frequency of 18.7~GHz the AME is detected with a significance of $21.2\sigma$, $10.2\sigma$ and $7.7\sigma$ respectively in W43r, W44r and W47r. The observed SEDs are fitted with a combination of free-free emission, a phenomenological model describing the spinning dust spectrum, and thermal dust emission, and in the case of W44r a synchrotron component, with $\chi_{\rm red}^2= 5.4$, 1.0 and 1.0 respectively in W43r, W44r and W47r, and $\chi_{\rm red}^2= 0.55$ in W44. The high value in W43r results from departures of the data from the model in the dust-dominated regime at frequencies $>70.4$~GHz. The fitted free-free amplitudes are in reasonable agreement with estimates derived from the RRL survey of \citet{alves12}, even in the SNR W44. On the other hand, the \commander component-separated maps from the \planck mission seem to overestimate the free-free emission and underestimate the AME contribution at higher frequencies. We find that the W44 intensity SED favours the presence of a free-free component, leading to a steeper synchrotron spectral index of $\beta_{\rm sync}=-0.72\pm 0.04$ than the values of $\approx -0.37$ found in previous studies \citep{castelletti07,green14,onic15,pip31} that consider a pure synchrotron power law. This value may have direct implications on the joint modelling of the radio and $\gamma$-ray data performed of \citet{cardillo14} and \citet{cardillo16}.

The inclusion of QUIJOTE data at eight frequencies between 10 and 20~GHz clearly helps to improve the modelling of the AME, decreasing the error bars in some parameters by a factor 6, while little affecting the other components. The shape of the resulting spinning dust spectra in W44r and in W47r look compatible with the theoretical predictions for WIM and WNM environments, whereas in the case of W43r the AME spectrum seems too broad. This shape could be explained by a mixture of different AME components contributing to the observed flux densities, something that seems possible given the complexity and richness of the W43 molecular complex. In all sources we get significant AME residual flux densities, reaching $269\pm 13$~Jy in the case of W43r, with AME emissivities (ratio between the AME peak and the $100~\mu$m flux density) in the range $(1.3-2.6)\times 10^{-4}$, consistent with what is usually found in HII regions, and a bit lower than in AME regions in general. Even when we try to better isolate the emission from the W44 source, we get a peak AME flux density of $40.3\pm 2.7$~Jy. This could be the first high-significance detection of AME originating in a SNR. A confirmation would require high-angular resolution observations in the range $10-30$~GHz, and also an assessment of the possible contribution from the nearby HII region G034.7-00.6, which is known to contain a PDR region dominated by PAHs, which are though to be the agents responsible for AME.

QUIJOTE and \wmap data exhibit strong polarisation associated with the synchrotron emission of W44, and significant diffuse emission distributed along the Galactic plane, that introduce some contamination around W43r and, to a higher extent, in W47r. We get at $22.7$~GHz a non-zero polarised flux density towards W43r, that could be associated with the diffuse synchrotron background, or even with a possible polarisation of the free-free emission. Neither its polarisation fraction nor the data at other frequencies allow us to disentangle between these two hypotheses. Even if we take this measurement as an upper limit, the polarisation fraction relative to the AME residual flux density is $\Pi_{\rm AME}<0.52\%$ (95\% C.L.) at this frequency, a factor 2 more stringent than previous best upper limits that were $\lesssim 1\%$ \citep{lopez11,dickinson11}. Except at \planck 28.4~GHz, which is clearly affected by systematics, and at DRAO $1.4$~GHz, where there seems to be a Faraday-screen effect of the background emission, we obtain polarisation consistent with zero at all other frequencies. At QUIJOTE 16.7~GHz, and \wmap $40.6$~GHz we get upper limits of $\Pi_{\rm AME}<0.39\%$ and $<0.22\%$. These constraints are inconsistent with all existing theoretical predictions for the polarisation fractions associated with magnetic dipole and electric dipole emissions, for the case of perfect alignment between the grain angular momentum and the magnetic field, for typical physical conditions and magnetic field strengths, and if we ignore the possible quantum dissipation of alignment \citep{draine16}. In particular, our constraints at 16.7~GHz, and between 30 and 45~GHz, are strongly inconsistent with the recent MD model of HL16 based on free-flying magnetic nanoparticles, for a grain temperature $T_{\rm d}=40$~K and grain size of 10~nm, which is the biggest size they consider. Even larger grains, or higher temperatures, that result in less efficient grain alignment, would be needed to bring the models in agreement with our data. In what concerns the ED models, our data below 45~GHz are inconsistent even with the HL16 model with the lowest magnetic field strength they consider, $5~\mu$G, and the highest temperature, $T_{\rm d}=60$~K. Therefore, in this case probably-unrealistic lower magnetic fields or higher temperatures would be needed to explain our data.

The polarisation SED of the W44 SNR is accurately modelled by a synchrotron power law with $\chi^2_{\rm red}=0.91$, when the DRAO $1.4$~GHz point, which seems affected by Faraday depolarisation, is not included in this fit. The fitted spectral index, $\beta_{\rm sync}=-0.62\pm 0.03$, is consistent with the one derived from the intensity SED when a free-free component is introduced in the model. This gives further support to the presence of free-free emission towards W44. We find a polarisation fraction relative to the residual synchrotron emission of $\Pi_{\rm sync}\sim 35\%$, which seems a bit high, but still possible in case of a highly-ordered magnetic field. Of course, when the free-free emission is excluded from the intensity fit, the modelled synchrotron intensity becomes higher, and in this case we get lower polarisation fractions of $\Pi_{\rm sync}\sim 10-15\%$. Finally, the change of the polarisation direction of W44 with the frequency, traced mainly by the QUIJOTE $10-20$~GHz data, indicates the presence of Faraday rotation. In our fit to a $\gamma=\gamma_0+{\rm RM}\times \lambda^2$ law we get $\chi_{\rm red}^2=1.00$ and a rotation measure of ${\rm RM}=-404\pm 49$~rad~m$^{-2}$, considerably larger than what is found by previous studies in this region.\\

\section*{Acknowledgements}
We thank Melis Irfan and Clive Dickinson for clarifying the procedure they used in \citet{irfan15} to obtain the flux densities in W43, W44 and W47. Bruce Draine and Thiem Hoang kindly provided us the predicted polarisation fractions associated with their models of magnetic and electric dipole emission presented in \citet{draine13} and \citet{hoang16}, respectively. Thanks are also given to Juan Betancort-Rijo, Sergio Colafrancesco, Eiichiro Komatsu, Carlos L\'opez-Caraballo and Juan Us\'on for providing enlightening comments. 

We acknowledge the use of data from the Planck/ESA mission, downloaded from the Planck Legacy Archive, and of the Legacy Archive for Microwave Background Data Analysis (LAMBDA). Support for LAMBDA is provided by the NASA Office of Space Science. Some of the results in this paper have been derived using the HEALP{\sc ix} \citep{gorski05} package. This work has been partially funded by the Spanish Ministry of Economy and Competitiveness (MINECO) under the projects AYA2007-68058-C03-01, AYA2010-21766-C03-02, AYA2012-39475-C02-01, the Consolider-Ingenio project CSD2010-00064  (EPI: Exploring the Physics of Inflation) and also by the European Union's Horizon 2020 research and innovation programme under grant agreement number 687312. FP thanks the European Commission under the Marie Sklodowska-Curie Actions within the H2020 program, Grant Agreement number: 658499-PolAME-H2020-MSCA-IF-2014.

\bibliographystyle{mn2e}

\appendix

\section{Correction factors}\label{ap:correction_factors}

In different sections of this article we have referred to some corrections that have been applied to the measured flux densities. In order to have a general overview, in Table~\ref{tab:correction_factors} we show the magnitude of these correction factors at each frequency and for each of the three studied regions. The first of these correction factors is used to transfer the flux scale of the \citet{reich86} and \citet{jonas98} maps from the full-beam to the main-beam, as it was explained in section~\ref{sec:ancillary_data}. In that same section we discussed that, in order to correct the contamination induced by the CO rotational transition lines on the \planck maps with frequencies 100, 217 and 353~GHz, we used the publicly-available Type 1 CO maps, which have the lowest systematic uncertainties \citep{cpp2013-13}. The procedure to apply the colour corrections was explained in section~\ref{subsec:intensity}. In that section we also justified the application of a correction factor to the C-BASS flux density for W44 taken from \citet{irfan15}, because they use different central coordinates. We have checked the reliability of the flux density adopted from C-BASS by verifying that this value is consistent (at $1.2\sigma$) with the prediction at the same frequency from a model derived from a fit to all frequencies excluding C-BASS.

\begin{table}
\begin{center}
\begin{tabular}{cccc}
\hline\hline\noalign{\smallskip}
Frequency &  W43r & W44r & W47r \\
(GHz) & (\%) & (\%) & (\%) \\
\noalign{\smallskip}\noalign{\smallskip}\hline\noalign{\smallskip}
 &\multicolumn{3}{c}{Full-beam to main-beam}\\
\noalign{\smallskip}\cline{2-4}\noalign{\smallskip}\noalign{\smallskip}
1.42   &    $55$     &    $55$     &    $55$     \\
2.33   &    $20$     &    $20$     &    $45$     \\
\noalign{\smallskip}\noalign{\smallskip}\hline\noalign{\smallskip}
 &\multicolumn{3}{c}{CO correction}\\
\noalign{\smallskip}\cline{2-4}\noalign{\smallskip}\noalign{\smallskip}
  100   &   $-27.5$   &   $-38.1$  &   $-29.0$  \\
  217   &   $-13.4$   &   $-16.4$  &   $-10.7$  \\
  353   &   $ -2.6$   &   $ -4.0$  &   $ -1.9$  \\
\noalign{\smallskip}\hline\noalign{\smallskip}
&\multicolumn{3}{c}{Colour corrections}\\
\noalign{\smallskip}\cline{2-4}\noalign{\smallskip}\noalign{\smallskip}
11.15   &  $  0.7$  &  $  0.6$   &   $  0.5$  \\
11.22   &  $  0.1$  &  $ -0.02$  &   $-0.04$  \\
12.84   &  $  0.5$  &  $  0.4$   &   $  0.4$  \\
12.89   &  $  0.5$  &  $  0.4$   &   $  0.4$  \\
16.75   &  $ 0.03$  &  $  0.1$   &   $  0.1$  \\
17.00   &  $  0.2$  &  $  0.2$   &   $  0.2$  \\
18.71   &  $  0.6$  &  $  0.6$   &   $  0.6$  \\
19.00   &  $  0.1$  &  $  0.1$   &   $  0.1$  \\
 22.7   &  $  0.6$  &  $  0.7$   &   $  0.7$  \\
 28.4   &  $  1.4$  &  $  1.1$   &   $  1.1$  \\
 32.9   &  $  0.3$  &  $  0.2$   &   $  0.2$  \\
 40.6   &  $  0.2$  &  $  0.1$   &   $ 0.07$  \\
 44.1   &  $  0.5$  &  $  0.3$   &   $  0.5$  \\
 60.5   &  $ -0.2$  &  $ -0.5$   &   $ -0.2$  \\
 70.4   &  $  1.2$  &  $  1.4$   &   $  1.7$  \\
 93.0   &  $ -1.7$  &  $ -2.1$   &   $ -1.8$  \\
  100   &  $ -5.3$  &  $ -6.2$   &   $ -5.5$  \\
  143   &  $ -1.8$  &  $ -1.9$   &   $ -1.9$  \\
  217   &  $-11.5$  &  $-11.5$   &   $-11.8$  \\
  353   &  $-11.2$  &  $-11.0$   &   $-11.4$  \\
  545   &  $-10.1$  &  $ -9.7$   &   $-10.1$  \\
  857   &  $ -2.5$  &  $ -2.1$   &   $ -2.3$  \\
 1249   &  $ -1.7$  &  $ -0.1$   &   $ -0.1$  \\
 2141   &  $  8.1$  &  $  7.0$   &   $  6.9$  \\
 2997   &  $  7.3$  &  $  8.7$   &   $  8.7$  \\
\noalign{\smallskip}\noalign{\smallskip}\hline\noalign{\smallskip}
 &\multicolumn{3}{c}{Coordinate offset}\\
\noalign{\smallskip}\cline{2-4}\noalign{\smallskip}\noalign{\smallskip}
4.76  &       $-$       &  $-12$  &   $-$  \\
\noalign{\smallskip}\hline\hline
\end{tabular}
\normalsize
\caption{Magnitude (percentage) of all the corrections that have been applied to the measured flux densities listed in Table~\ref{tab:fluxes}.}
\label{tab:correction_factors}
\end{center}
\end{table}

\section{Systematic uncertainties}\label{ap:systematics}

One of the main sources of systematic uncertainties are the calibration errors, which have been disregarded in the analyses presented in this article. These uncertainties, as given in the corresponding references, are listed in Table~\ref{tab:fluxes}. We have also added: i) a 5\% relative error to the \citet{reich86} and \citet{jonas98} maps associated with the full-beam to main-beam correction factors discussed in section~\ref{sec:ancillary_data}; ii) a 5\% relative error to the C-BASS flux densities due to the scaling factor to account for the difference between \citet{irfan15} and our coordinates (see section~\ref{subsec:intensity}); iii) the calibration errors of the Type 1 CO maps \citep{cpp2013-13}, that are $10\%$, $2\%$ and $5\%$ respectively for 100, 217 and 353~GHz. 

We used Type 1 maps because they have smaller systematic uncertainties than Type 2 and Type 3 maps, which have also been made publicly-available by the \planck collaboration \citep{cpp2013-13}. The level of the resulting corrections is shown in Table~\ref{tab:correction_factors}. Note that the previous CO calibration errors are referred to the flux density associated with the CO emission and not to the total measured flux density. The actual errors associated with the total measured flux densities are therefore smaller. Whereas we use these data points in our fits, in other analyses \citep{per20,pip15,irfan15} the 100 and 217~GHz points are removed in order to avoid possible residual CO contamination, while the 353~GHz, which is less severely affected, is kept in but with no correction applied. In the Table~\ref{tab:parameters_noco} we show the resulting best-fit parameters when the 100, 217 and 353~GHz are excluded from the fit. A comparison with the parameters shown in Table~\ref{tab:parameters}, when all frequencies are included, shows that the differences are typically below 3\%. Further to this, we have also checked that the flux densities predicted at 100, 217 and 353~GHz by this fitted model are always consistent within $1\sigma$ with those derived from the CO-corrected maps. This demonstrates the reliability of the CO correction that we have applied, which seems not to introduce systematic errors at any level higher than our statistical error bars.

\begin{table}
\begin{center}
\begin{tabular}{lccc}
\hline\hline
\noalign{\smallskip}
  & W43r & W44r  & W47r \\
\noalign{\smallskip}
\hline
\noalign{\smallskip}
                      $EM$  (cm$^{-6}$~pc)   &   3907  $\pm$	29  &	1194  $\pm$    22  &   1850  $\pm$    20   \\
\noalign{\smallskip}
 $S_{\rm sync}^{\rm 1 GHz}$ (Jy)             &       --             &	 230  $\pm$	7  &	  --		   \\
\noalign{\smallskip}
              $\beta_{\rm sync}$             &       --             &  -0.59  $\pm$  0.04  &	  --		   \\
\noalign{\smallskip}
                        $m_{60}$             &   1.58  $\pm$  0.16  &	3.26  $\pm$  1.30  &   5.20  $\pm$  1.41   \\
\noalign{\smallskip}
$\nu_{\rm AME}^{\rm peak}$ (GHz)             &   22.2  $\pm$   0.5  &	21.5  $\pm$   1.0  &   20.7  $\pm$   0.9   \\
\noalign{\smallskip}
   $S_{\rm AME}^{\rm peak}$ (Jy)             &  258.2  $\pm$   3.0  &	79.7  $\pm$   5.7  &   42.7  $\pm$   2.3   \\
\noalign{\smallskip}
              $\beta_{\rm dust}$             &   1.77  $\pm$  0.03  &	1.76  $\pm$  0.04  &   1.89  $\pm$  0.05   \\
\noalign{\smallskip}
                  $T_{\rm dust}$  (K)        &   21.9  $\pm$   0.3  &	20.0  $\pm$   0.5  &   19.5  $\pm$   0.4   \\
\noalign{\smallskip}
                    $\tau_{250}$             &   4.26  $\pm$  0.23  &	2.30  $\pm$  0.22  &   2.57  $\pm$  0.25   \\
\noalign{\smallskip}
              $\chi^2_{\rm red}$             &        5.5	    &	     1.2	   &	    1.1	    \\
\noalign{\smallskip}
\hline\hline
\end{tabular}
\normalsize
\caption{Best-fit model parameters for the three regions, when the 100, 217 and 353~GHz \planck data points, which are potentially contaminated by CO emission, are not used in the fit. In the last line we show the reduced chi-squared.}
\label{tab:parameters_noco}
\end{center}
\end{table}

Another possible systematic effect could be zodiacal light emission. As indicated in section~\ref{sec:ancillary_data} we used zodi-subtracted COBE-DIRBE maps. For {\it Planck}-HFI frequency bands there are also publicly-available zodi-corrected maps, although we have used the uncorrected ones. However, thanks to the smoothness of the zodiacal light emission, our background subtraction leads to a severe cancellation of this emission. In fact, we have re-computed flux densities using the zodi-corrected {\it Planck}-HFI maps and found that the differences are always below 0.02\%.

In their analyses \citet{irfan15} assign an additional 3\% systematic uncertainty associated with beam asymmetries in the \wmap and \planck data. This value seems to have been chosen in an arbitrary way. A more accurate value of the error introduced by this effect could be assessed through simulations, but that is beyond the scope of this article. However, we note that the smoothing procedure that has been followed to take all the maps to a common angular resolution of $1^\circ$ reduces the impact of beam asymmetries in the determination of our flux densities.

\section{Error treatment}\label{ap:errors}

As it was explained in section~\ref{sec:intensity}, the error bars associated with our flux density estimates (Table~\ref{tab:fluxes}) have been determined through the RMS dispersion of the data in the background annulus, and therefore represent statistical uncertainties only. Some authors \citep{per20,irfan15} choose to combine the statistical and systematic errors in quadrature. This would be a good approach if the systematic errors were uncorrelated. However, we know that the calibration uncertainties of different frequency bands of the same experiment will be correlated to some degree, and therefore would tend to shift the measurement in the same direction. For this reason, ideally the inclusion of the systematic errors when fitting models to the measured SEDs would require the knowledge of their covariance matrix. As we lack this information we decided to use only statistical errors. Even so, as a sanity check, we have repeated the fits after combining the statistical and systematic errors in quadrature. The resulting models are in practice undistinguishable, as it becomes evident when we compare the best-fit parameters of Table~\ref{tab:parameters_sys} with those of Table~\ref{tab:parameters}. The change in most of the parameters is very small, typically below $5\%$. In particular, the amplitude of the AME changes only by $-0.5$\%, $-3.4$\% and $+0.1$\%, respectively in W43r, W44r and W47r. The biggest change is in $\tau_{250}$, as a consequence of the large calibration uncertainties of the DIRBE data.

\begin{table}
\begin{center}
\begin{tabular}{lccc}
\hline\hline
\noalign{\smallskip}
  & W43r & W44r  & W47r \\
\noalign{\smallskip}
\hline
\noalign{\smallskip}
                            $EM$  &   3877  $\pm$    54  &   1357  $\pm$    32  &  1838  $\pm$    28   \\
\noalign{\smallskip}
 $S_{\rm sync}^{\rm 1 GHz}$       &	 --		 &    205  $\pm$    16  &     --	       \\
\noalign{\smallskip}
              $\beta_{\rm sync}$  &	 --		 &  -0.66  $\pm$  0.08  &     --	       \\
\noalign{\smallskip}
                        $m_{60}$  &   1.44  $\pm$  0.26  &   3.16  $\pm$  1.66  &  5.00  $\pm$  1.94   \\
\noalign{\smallskip}
$\nu_{\rm AME}^{\rm peak}$        &   22.0  $\pm$   1.0  &   21.0  $\pm$   1.4  &  20.7  $\pm$   1.2   \\
\noalign{\smallskip}
   $S_{\rm AME}^{\rm peak}$       &  256.8  $\pm$   5.5  &   75.1  $\pm$   8.9  &  42.8  $\pm$   3.5   \\
\noalign{\smallskip}
              $\beta_{\rm dust}$  &   1.64  $\pm$  0.04  &   1.73  $\pm$  0.06  &  1.82  $\pm$  0.06   \\
\noalign{\smallskip}
                  $T_{\rm dust}$  &   24.1  $\pm$   0.8  &   20.4  $\pm$   0.8  &  20.3  $\pm$   0.8   \\
\noalign{\smallskip}
                    $\tau_{250}$  &   2.96  $\pm$  0.32  &   2.10  $\pm$  0.29  &  2.17  $\pm$  0.31   \\
\noalign{\smallskip}
              $\chi^2_{\rm red}$  &	  0.9  	 &	 0.3    	&	 0.3	       \\
\noalign{\smallskip}
\hline\hline
\end{tabular}
\normalsize
\caption{Best-fit model parameters for the three regions, obtained when we add in quadrature the systematic uncertainties to the statistical errors quoted in Table~\ref{tab:fluxes}. In the last line we show the reduced chi-squared}
\label{tab:parameters_sys}
\end{center}
\end{table}

Judging by the reduced chi-squared values of Table~\ref{tab:parameters_sys}, the addition of the calibration uncertainties leads to an overestimation of the errors. This might be a consequence of having disregarded the correlations between the calibration errors of different bands of the same experiment. On the other hand, our statistical errors might also be slightly overestimated, as we have not taken into account in the calculation of the RMS that the noise component due to background fluctuations is correlated between individual pixels. In fact, in regions close to the Galactic plane and at angular scales of $\sim 1^\circ$ background fluctuations might be the dominant contribution to the error bar rather than instrument noise. A perfect characterisation of the noise would require in this case the knowledge of the full noise covariance matrix, including separately the contributions from instrument and background noise. Given our lack of knowledge of the background correlation function, an alternative way of calculating the errors is to define a set of random positions around the source, and to calculate the dispersion of the flux densities derived in those positions, using the same sizes for the aperture and for the background annulus. The result of this analysis, performed in 10 positions with Galactic latitudes in the range $2^\circ-10^\circ$, is shown in Table~\ref{tab:err_fluxes_ra}, in comparison with the errors shown in Table~\ref{tab:fluxes}, which were based on the pixel-to-pixel RMS dispersion calculated in the background annulus. The random apertures lead to values that are of the same order but generally lower, particularly at frequencies above 10~GHz. The reason of being in general lower is probably that we capture in our background annulus a large dispersion produced by the gradient of the Galactic emission, something that does not happen in apertures located away from the Galactic plane. In this respect, our assumption that the noise RMS in the aperture will be the same as in the background annulus must be regarded as conservative, as the aperture extends to smaller distances from the Galactic plane and therefore will have a smaller dispersion associated with the gradient of the Galactic emission.

\begin{table}
\begin{center}
\begin{tabular}{ccccc}
\hline\hline
\noalign{\smallskip}
Freq. & \multicolumn{4}{c}{Error on the flux density (Jy)} \\
\noalign{\smallskip}
\cline{2-5}
\noalign{\smallskip}
(GHz) & Random & \multicolumn{3}{c}{Pixel-to-pixel RMS}\\
 & apertures & W43r & W44r & W47r\\
\noalign{\smallskip}
\hline
\noalign{\smallskip}
      0.408     &        24     &      21    &      23   &       22  \\
      0.820     &        16     &      18    &      15   &       15  \\ 
      1.42      &        12     &      17    &      12   &       10  \\
      2.32      &        14     &      18    &      16   &       14  \\
      11.15     &         5     &       9    &       7   &        5  \\
      11.22     &         5     &       6    &       6   &        5  \\
      12.84     &         6     &       7    &       6   &        6  \\
      12.89     &         3     &      10    &       7   &        6  \\
      16.75     &         4     &      10    &       7   &        6  \\
      17.00     &         6     &       9    &       6   &        6  \\
      18.71     &         5     &      11    &       7   &        6  \\
      19.00     &         7     &      10    &       7   &        6  \\
      22.7      &         4     &      15    &      10   &        9  \\
      28.4      &         4     &      15    &       9   &        8  \\
      32.9      &         4     &      14    &       9   &        8  \\
      40.6      &         4     &      13    &       8   &        7  \\
      44.1      &         4     &      12    &       8   &        6  \\
      60.5      &         4     &      11    &       7   &        6  \\
      70.4      &         5     &      12    &       8   &        7  \\
      93.0      &         8     &      16    &      12   &       12  \\
      100       &        10     &      18    &      14   &       14  \\
      143       &        23     &      44    &      37   &       41  \\
      217       &        94     &     172    &     149   &      165  \\
      353       &       460     &     836    &     728   &      797  \\
      545       &      1646     &    3257    &    2749   &     2936  \\
      857       &      4299     &   12666    &   10264   &    10490  \\
      1249      &      6940     &   31268    &   24490   &    23547  \\
      2141      &     10553     &   59959    &   41812   &    36245  \\
      2997      &      8388     &   31289    &   20297   &    15778  \\
\noalign{\smallskip}
\hline\hline
\end{tabular}
\normalsize
\caption{Comparison between the estimates of the error on the flux density estimates, obtained from random apertures around the source, and from the pixel-to-pixel dispersion on the background annulus around each source.}
\label{tab:err_fluxes_ra}
\end{center}
\end{table}

In summary, it is clear that a better assessment of the error bars is not possible due to the lack of knowledge of the true covariance matrix of the noise. However, we have shown compelling evidence that we have provided sufficiently reliable and conservative error bars. It must also be noted that we normalise the error bars of the best-fit parameters using $\sqrt{\chi^2_{\rm red}}$, so they would essentially be very similar independently on the inclusion or not of the calibration errors.

\pagestyle{plain}

\bsp
\label{lastpage}
\end{document}